\documentclass[manuscript,screen,nonacm]{acmart}
\AtBeginDocument{%
  }

\setcopyright{acmlicensed}
\copyrightyear{2026}
\acmYear{2026}
\acmDOI{XXXXXXX.XXXXXXX}
\acmConference[Conference acronym 'XX]{Make sure to enter the correct
  conference title from your rights confirmation email}{June 03--05,
  2018}{Woodstock, NY}
\acmISBN{978-1-4503-XXXX-X/2018/06}

\usepackage{graphicx}
\usepackage{multirow}
\usepackage{caption,subcaption,graphicx}
\usepackage{xcolor}

\begin{document}

\title{RuleEdit: Failure-Guided Human-AI Model Editing with Prospective Impact Preview}

\author{Min Hun Lee}
\email{mhlee@smu.edu.sg}
\author{Justin Yu Feng Teo}
\email{justin.teo.2022@scis.smu.edu.sg}
\affiliation{%
  \institution{Singapore Management University}
  \city{Singapore}
  \country{Singapore}
}

\renewcommand{\shortauthors}{Lee and Teo}

\begin{abstract}
Despite the promise of AI to assist complex decisions, practitioners still lack ways to detect likely failures and inspect the consequences of model edits before committing them. We present RuleEdit, an interactive, rule-guided human-AI model editing system that (i) surfaces likely failures through interpretable mismatch signals from rule tables and (ii) supports user-authored rule feedback with prospective previews of projected performance changes and embedding shifts. We instantiate RuleEdit in stroke rehabilitation assessment and evaluate it with health professionals and students. Rule-guided failure detection significantly increased Human + AI performance by 14.16\% ($p<0.001$) while improving rejection of incorrect AI and reducing both over- and under- reliance as well as ChangedToWrong decisions. In addition, presenting prospective embedding previews improved participants' feedback for model adaptation, increasing post-update local performance gains from 11.50\% to 36.38\% after incorporating users' rule-based feedback ($p<0.001$). Our findings show that mismatch-based failure cues and prospective impact previews can support failure-aware human-AI model editing, while also revealing a local-global tradeoff: edits that help a specific case can degrade performance when transferred globally. We discuss implications of designing failure-aware and controllable human-AI systems.

\end{abstract}

\maketitle

\section{Introduction}
Human–AI collaborative decision-making \cite{lai2021towards,lee2021human,salimzadeh2024dealing,cai2019hello,cai2019human,holstein2023toward,buccinca2021trust} examines how humans and AI systems can combine complementary strengths to improve decisions in high-stakes domains such as healthcare \cite{cai2019human,lee2021human,kim2024much}, law \cite{lima2021human}, and public services \cite{de2020case,stapleton2022imagining,kawakami2022improving,kuo2023understanding}. To support end users, prior work has investigated data-driven scaffolds including explanations that highlight important features \cite{lee2021human,wang2021explanations}, retrieve similar cases \cite{cai2019human,chen2023understanding}, counterfactuals \cite{lee2023understanding,zhang2022towards}, and uncertainty cues \cite{zhang2020effect,prabhudesai2023understanding,li2025confidence,lee2025towards}. These mechanisms aim to clarify why a model produced a given output \cite{arya2019one,wang2019designing,lakkaraju2020explaining,abdul2018trends,preece2018asking} and to quantify confidence \cite{zhang2020effect,prabhudesai2023understanding,li2025confidence,lee2025towards} so that people can judge when to rely on AI. 

While explanations and confidence cues can improve users’ mental models of AI \cite{cai2019human,lee2021human,beede2020human,wang2021brilliant,lee2023understanding,prabhudesai2023understanding}, some studies show that explanations can inflate unwarranted trust and induce over-reliance on erroneous outputs \cite{bussone2015role,bansal2021does,buccinca2021trust,lee2023understanding}. Achieving calibrated reliance and human-AI complementarity \cite{lai2021towards,bansal2021does,holstein2023toward,buccinca2021trust} remains challenging. 
A core gap remains: most systems emphasize informational transparency showing explanations and scores, but provide limited support for two practical tasks. First, users often lack actionable cues for identifying when the AI is likely wrong on a specific case. Second, even when users want to intervene, they rarely receive previews of how a proposed edit may affect model behavior before it is committed.

\begin{figure*}[htp]
\centering 
\includegraphics[width=1.0\linewidth]%
  {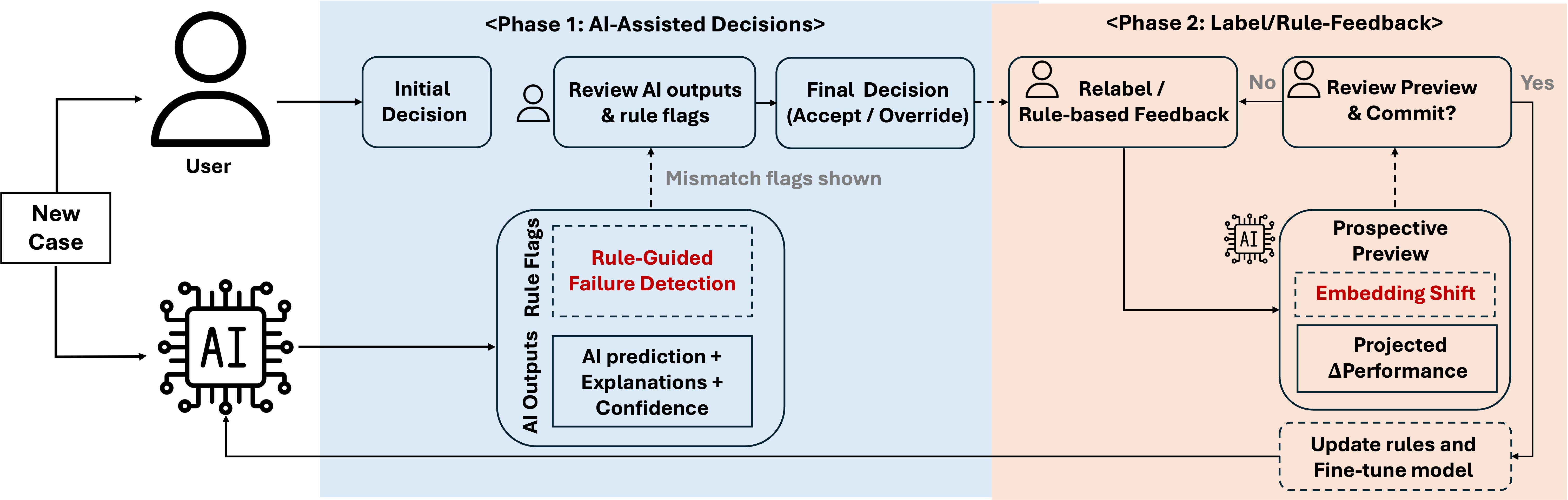}
\caption{RuleEdit supports failure inspection, user-authored editing, and prospective preview in AI-assisted decision-making. Given a new case, users first make an initial judgment and then review the system’s AI output together with rule-guided mismatch cues that highlight likely inconsistencies between the model prediction and rule-based evidence. Users then finalize their decision (accept/override). For refinement, users can provide label feedback (relabel) and/or rule feedback (select rules and edit thresholds). The system computes a prospective preview including projected $\Delta$ performance and embedding-shift summaries before users decide whether to commit the update.}\label{fig:flow_diagram}
\end{figure*}

\begin{figure*}[htp]
\centering 
\begin{subfigure}[t]{0.5\textwidth}
  \centering
\includegraphics[width=1.0\columnwidth]%
  {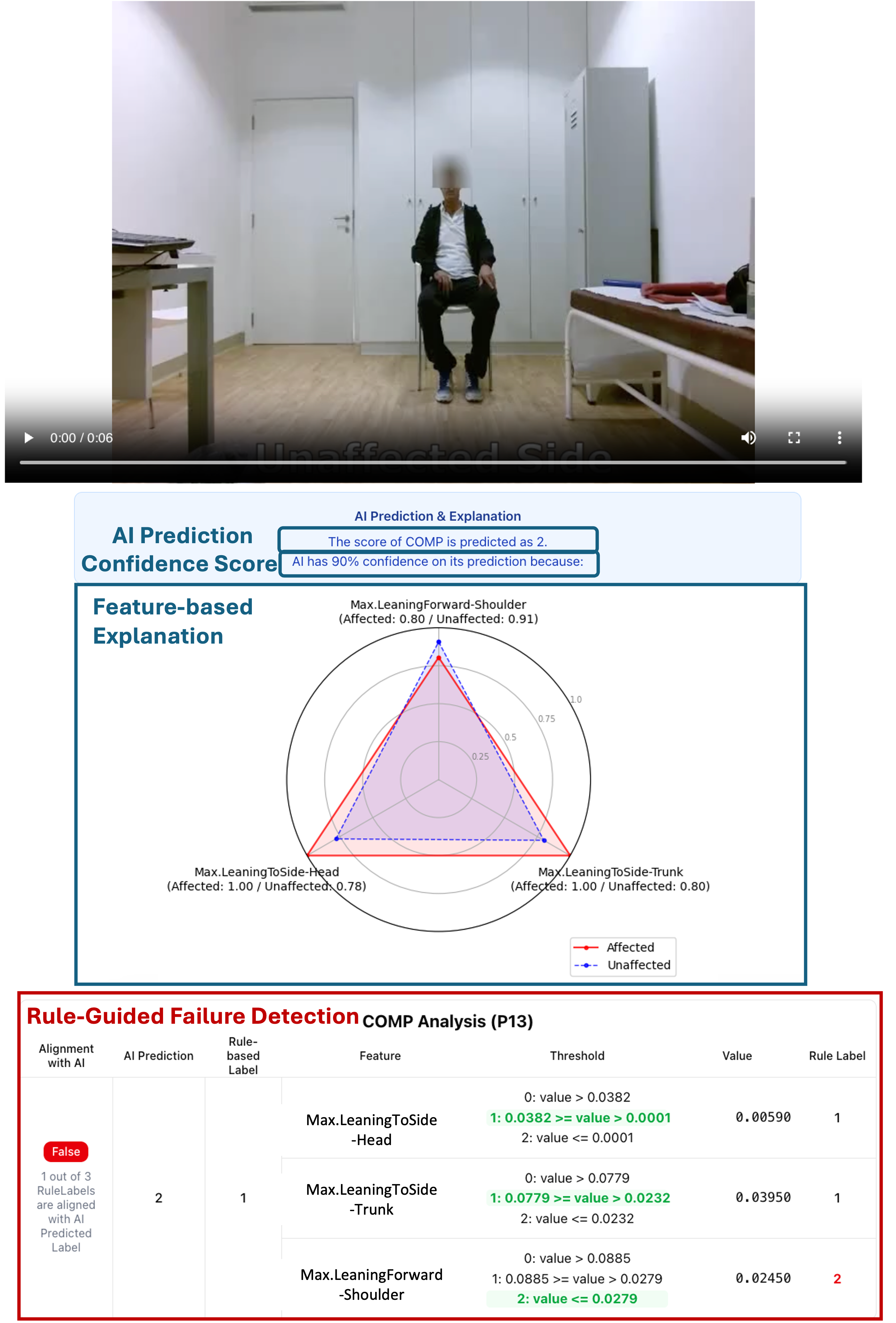}
  \caption{}
  \label{fig:interface_phase1}
\end{subfigure}
\begin{subfigure}[t]{0.49\textwidth}
  \centering
\includegraphics[width=1.0\columnwidth]%
  {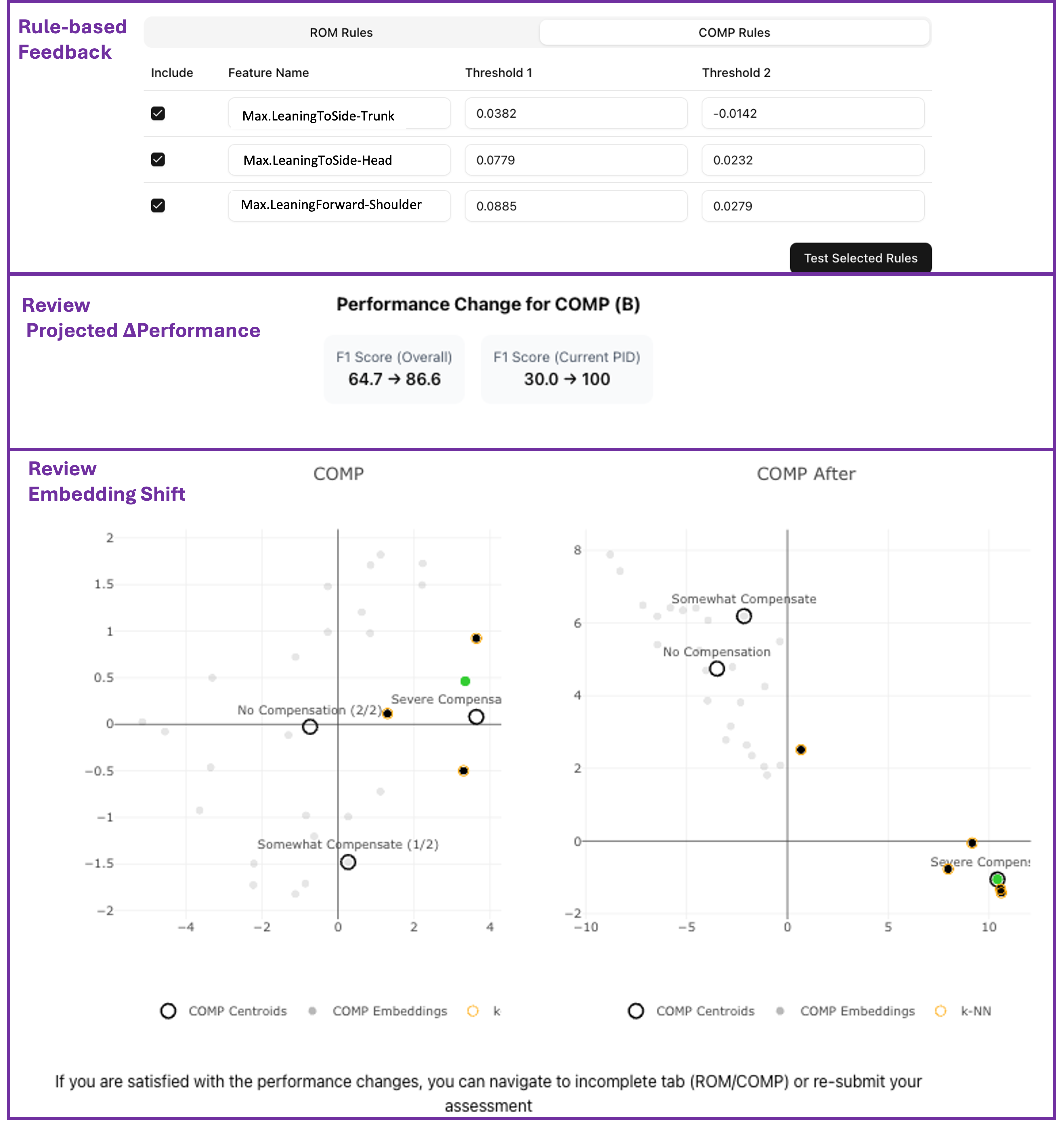}
  \caption{}
  \label{fig:interface_phase2}
\end{subfigure}
\caption{The system interface includes the video of a post-stroke survivor and the corresponding AI output (prediction, confidence score, and feature-based explanation). (a) The failure-inspection view, where mismatch cues surface likely disagreements between the AI prediction and rule-based evidence to support accept/override decisions, and (b) the editing view, where users select or edit rules and inspect projected performance change together with before/after embedding-shift previews (updated neighborhoods and class centroids) before committing an update.}\label{fig:interface}
\end{figure*}

To address this gap, we present RuleEdit, an interactive, failure-guided human-AI model editing system that surfaces likely failures through interpretable mismatch signals derived from rule tables and supports user-authored rule and label feedback for model editing. Beyond helping users inspect failures after an AI prediction is shown, RuleEdit supports anticipatory interaction by previewing projected performance changes and embedding shifts (on class centroids and data) before users decide whether to commit an edit. 

We instantiate RuleEdit in physical stroke rehabilitation assessment and conduct a study with 21 participants including health professionals and students to investigate two research questions: (RQ1) How do rule-guided mismatch cues affect users' accept/override decisions and Human + AI team performance? and (RQ2) How does prospective impact preview affect the quality of user-authored feedback for model editing?

To explore these research questions, we use a dataset of 15 post-stroke survivors and 10 healthy participants \cite{lee2019learning,hun2023design} and developed an AI-based decision-support system with a feed-forward neural network to classify post-stroke patients' quality of motion. Grounded in prior findings that health professionals value interpretable, important feature explanations \cite{lee2021human}, our system also presents feature-importance explanations to assist users' rehabilitation assessment tasks (Figure \ref{fig:interface_phase1}). 
To support AI failure detection, we implemented a rule-based table that flags case-level contradictions between rule verdicts and model predictions (Figure \ref{fig:interface_phase1}). For this rule-based table, we empirically selected the top three motion-quality rules by validation performance. To preview prospective embedding updates, showing how data points and class centroids would shift after applying rule feedback, we utilize supervised Uniform Manifold Approximation and Projection (UMAP) \cite{becht2019dimensionality} with supervision given by user-provided relabels on disagreed cases and updated AI predictions from an ensemble approach that integrates rule-based feedback while keeping the underlying ML model frozen.

To evaluate the effectiveness of our rule-guided failure detection and model adaptation with prospective performance estimates and prospective embedding previews, we conducted a within-subjects experiment with domain experts (i.e. therapists with experience in stroke rehabilitation) and novices (i.e. other health professionals with limited experience in stroke rehabilitation and students in medicine, nursing, and therapy). The study separates two questions: whether rule-guided mismatch cues improve AI-assisted decision-making at decision time, and whether prospective impact preview improves the quality of user-authored edits at refinement time.  

Our study shows that rule-guided failure detection increased Human+AI team performance by 14.16\% ($p<0.001$) and improved failure identification while reducing both over- and under-reliance and  ChangedToWrong decisions by 20.53\% ($p<0.01$), 7.15\% ($p<0.01$), and 8.00\% ($p<0.01$) respectively. In addition, presenting prospective impact previews improved the quality of user-authored model edits, raising post-update local performance gains from 11.50\% to 36.38\% ($p<0.001$) after incorporating users’ rule-based feedback. At the same time, our results reveal a local--global tradeoff: edits that improve a specific case locally may not transfer safely when pooled into broader global updates.

Overall, our work contributes to research on human-AI collaborative decision-making by (i) presenting RuleEdit, an interactive, failure-guided human-AI model editing system that combines mismatch-based failure inspection with user-authored edits and prospective impact preview; (ii) showing that rule-guided mismatch cues can improve calibrated reliance and Human+AI team performance during AI-assisted decision-making; and (iii) demonstrating that prospective preview supports higher-quality local refinement while exposing an important boundary: naively pooled global updates can regress despite strong local gains. These findings motivate safeguards for failure-aware and controllable human-AI systems.

\section{Related Work}
\subsection{Human-AI Collaborative Decision-Making}
Human-Artificial Intelligence (AI) collaborative decision making research \cite{lai2021towards,lee2021human,salimzadeh2024dealing,cai2019hello,cai2019human,holstein2023toward,buccinca2021trust} investigates how humans and AI can leverage complementary strengths to improve decision-making in various fields (e.g. healthcare \cite{cai2019human,lee2021human,kim2024much}, law \cite{lima2021human}, and other public services \cite{de2020case,stapleton2022imagining,kawakami2022improving,kuo2023understanding}). To support human decision-making, prior work provides decision-support aids, such as feature-based explanations \cite{lee2021human,wang2021explanations}, similar-case retrieval \cite{cai2019human,chen2023understanding,lee2024interactive}, counterfactual explanations \cite{lee2023understanding,zhang2022towards} alongside uncertainty cues  \cite{zhang2020effect,prabhudesai2023understanding,li2025confidence,lee2025towards,kim2024m} and model documentation (e.g., model cards) \cite{mitchell2019model,lee2024improving}. These approaches aim to help users interpret why an AI model produced a prediction \cite{arya2019one,wang2019designing,lakkaraju2020explaining,abdul2018trends,preece2018asking}, estimate when that prediction may be uncertain \cite{zhang2020effect,prabhudesai2023understanding,li2025confidence,lee2025towards}, or communicate system-level performance characteristics \cite{mitchell2019model} to support users to determine when to trust AI. 

These techniques can improve humans’ accuracy and efficiency in practice \cite{cai2019human,lee2021human,beede2020human,wang2021brilliant,lee2023understanding,prabhudesai2023understanding}. However, findings on their effectiveness are mixed. While some studies find limited effects of explanations on trust \cite{cheng2019explaining} or improve decision quality \cite{zhang2020effect,lee2021human,prabhudesai2023understanding,lai2019human}, other studies show that AI explanations could inflate trust in AI and induce overreliance on incorrect AI outputs \cite{bussone2015role,bansal2021does,buccinca2021trust,lee2023understanding}. Fostering calibrated reliance remains challenging as over-trust propagates errors  and undermines human-AI complementarity \cite{lai2021towards,bansal2021does,holstein2023toward,buccinca2021trust}. Much of this work helps users interpret AI outputs, but offers limited support for two practical tasks: identifying likely failures on a specific case \cite{wu2019errudite,cabrera2021discovering,tang2024ai} and intervening in a way that makes the consequences of an edit visible before it is applied.

Our work builds on this literature by treating human-AI decision-making not only as an interpretive task, but also an editing task: users must inspect likely failures, decide whether to accept or override the AI, and, when needed, propose refinements to system behavior. RuleEdit contributes an approach that supports these actions through mismatch-based failure cues and prospective preview before commit for trustworthy human-AI collaboration.

\subsection{Interactive Failure Detection and Debugging}
Moving beyond presenting static explanations, researchers have developed interactive approaches that let users probe, test, and refine model behavior. Example-based views, counterfactual exploration, and what-if analysis help people interrogate decision boundaries and inspect failure modes \cite{kulesza2015principles,lakkaraju2016interpretable,lee2023understanding}. Interactive debugging paradigms further enable users to flag errors, adjust system behavior, and provide feedback that influences subsequent predictions \cite{amershi2014power,yang2018grounding,teso2019explanatory,kulesza2015principles}. Within computational interaction, carefully designed feedback channels, such as labeling and co-auditing, have been shown to support oversight and iterative improvement \cite{kulesza2015principles,amershi2014power,cheng2019explaining}. 

A central challenge in these systems is making the path from user feedback to model change understandable and trustworthy. Users need not only mechanisms to provide corrections, but also ways to anticipate what a proposed intervention will do before it is committed. Prior work investigates various approaches for interactive debugging from re-labeling 
\cite{lee2021pebble} 
and allocating human effort where uncertainty is high \cite{lee2025towards,wilder2020learning,raghu2019transfusion}, to manipulating confusion matrix to steer a model \cite{kapoor2010interactive}, and adjusting feature weights \cite{kulesza2011oriented} or feature relevance \cite{kulesza2015principles,lee2021human,guo2022building}. However, most prior systems emphasize retrospective correction: users identify an issue, provide feedback, and only then observe updated behavior. Support for anticipatory intervention, where users preview the likely downstream effects of an edit before commit, remains limited. 

RuleEdit extends this line of work by making candidate edits legible before commitment. The system previews projected performance changes together with representation shifts, such as embedding neighborhoods \cite{boggust2022embedding} and class centroids, under a proposed update. These previews support anticipatory auditing by helping users inspect whether an intervention is likely to improve behavior locally or introduce broader regressions.

\subsection{Rule-Based Representations for Interpretability and Control
}
Rules provide a structured, interpretable interface for users to interact with AI systems. Unlike open-ended or example-based feedback \cite{dudley2018review}, rules establish explicit, auditable links between human understanding and model behavior, enabling practitioners to see when and how their inputs will take effect \cite{rudin2019stop,ribeiro2018anchors}. Prior work on interpretable decision sets and human-in-the-loop approaches shows that such structured representations can both explain predictions and constrain them at inference or training time \cite{lakkaraju2016interpretable,ribeiro2018anchors,kulesza2015principles,lee2021human,guo2022building,frisch2005rules}.

Rule-based approaches have long served as a way to externalize and revise decision logic \cite{shortliffe1986medical}. In contemporary AI systems, they remain attractive as interaction primitives because they can capture domain constraints, encode safety conditions, and provide users with an interpretable handle for influencing model behavior \cite{rudin2022interpretable,green2019principles}. However, we still know relatively little about how to design rule-guided workflows that are usable by non-ML experts during real decision tasks, especially when those workflows must support both failure inspection and safe refinement.

We address this gap with a rule-guided editing system in which users leverage interpretable rules to (i) surface likely failure conditions and (ii) propose controllable adaptations. In contrast to static explanations and uncertainty displays \cite{lee2021human,cai2019human,wang2021explanations,zhang2020effect,prabhudesai2023understanding,lee2023understanding,li2025confidence,lee2025towards}, our approach makes candidate interventions explicit and previews their likely effects before commitment. We empirically examine how these interactions shape human-AI collaborative decision-making for rehabilitation assessment with health professionals and health-related students.

\section{Study Design}
This work investigates whether interpretable rules can assist users in detecting AI failures and enacting controllable adaptations for human-AI collaborative decision-making (i.e. physical stroke rehabilitation assessment). Building on prior work that explores the utility of informational aids \cite{buccinca2021trust,lee2023understanding,prabhudesai2023understanding}, we examine two interaction questions: whether rules can (i) surface likely failure cases in situ to improve how users rely on or override the model during decision-making and (ii) support user-authored model refinements through pre-commit preview. To make these refinements more understandable and auditable, we examine the effect of showing projected performance change together with updated representation views including embedding points and class centroids, before an edit is committed. This framing emphasizes anticipatory rather than purely reactive interaction.

To study these questions, we proceed in two stages. First, we evaluate how well rule-derived signals identify likely AI failures relative to the model's predictions. Second, we conduct a user study with health professionals and medical/health students to examine two interface effects: whether rule-guided mismatch cues improve accept/override decisions during AI-assisted assessment, and whether pre-commit preview improves the quality of user-authored edits during model refinement.

\subsection{Study Context: Physical Stroke Rehabilitation Assessment}
We situate our study in a clinical decision-making task: assessing the movement quality of post-stroke survivors. Following prior work on human–AI collaboration for this domain \cite{lee2021human}, we focus on a functional task in which participants raise the wrist to the mouth, approximating a drinking motion. For rehabilitation assessment, we consider two commonly used dimensions: Range of Motion (ROM) and compensation \cite{lee2020co}. ROM captures how completely the target posture is achieved (i.e. whether the hand reaches the intended position). Compensation captures the use of unintended strategies to accomplish the task, such as shoulder elevation or trunk leaning, indicating reliance on non-target joints.

\subsection{System Implementations}
Our system for AI-assisted stroke rehabilitation assessment combines a feed-forward neural network classifier with two interaction components: rule-guided failure inspection at decision time and user-authored editing with prospective preview at refinement time (Figure \ref{fig:interface}). The system presents each rehabilitation video alongside the AI output (prediction, confidence, and feature-based explanation) and mismatch-based failure cues that surface likely inconsistencies between the model prediction and rule-based evidence. Users make an initial decision, review the AI output and mismatch cues, and then finalize the decision by accepting or overriding the AI prediction (Figure \ref{fig:flow_diagram}). For interactive refinement (Figure \ref{fig:interface_phase2}), the system supports label feedback (i.e. re-label disagreed cases) and rule-based feedback (i.e. select/edit rules and adjust thresholds). Before committing an update, an on-demand, interactive preview module estimates the expected impact on prospective $\Delta$ performance change and prospective embedding-shift (updated neighborhoods and class centroids) while keeping the underlying NN frozen. The interface is implemented using a client–server architecture: a React frontend renders the study flow and interactive views, while a Python/Flask \cite{grinberg2018flask} backend with PostgreSQL (SQLAlchemy) supports session-based APIs and stores user study data.

\subsubsection{Dataset}
We draw on an upper-limb rehabilitation dataset on the “bring a cup to the mouth” task. The dataset comprises (i) 300 videos from 15 post-stroke survivors with varied motor impairments,  including the affected and unaffected side,  \cite{lee2019learning} and 60 videos from 10 healthy participants who performed one correct motion and five acted impairments designed to mimic common post-stroke movement errors \cite{hun2023design}, (ii) 3D skeletal joint trajectories extracted using a Microsoft Kinect v2, and (iii) ground truth labels by expert therapists conducted a clinically validated assessment (e.g. Fugl–Meyer Assessment \cite{gladstone2002fugl}) at recruitment.

\subsubsection{AI Model}\label{sect:ai_model}
Following prior work on rehabilitation assessment \cite{lee2019learning}, we extracted a set of kinematic features of each trial (Appendix~\ref{appendix:feature_extract}). As feed-forward neural network (NN) models perform competitively for assessing quality of rehabilitation exercises \cite{lee2019learning}, we implemented feed-forward NN models in PyTorch \cite{paszke2019pytorch} for Range of Motion (ROM) and Compensation. For the ROM, the model utilizes three hidden layers of 64 units each with a learning rate of 1e-2. For Compensation, the model uses three hidden layers (32, 32, 64 units) with a learning rate of 5e-3. Using leave-one-subject-out (LOSO) cross-validation, the models achieved average F1 scores of 82.83\% for ROM and 80.74\% for Compensation.

\subsubsection{Rule-based Model and Hybrid Inference}\label{section:rb_model}
We implemented a family of lightweight rule-based models \cite{varshney2023literature} to map each kinematic feature to a clinical label ($y \in \{0, 1, 2\}$) for each performance element (i.e. ROM and COMP). For a given element and feature $f$, we learn a pair of thresholds ($T_1, T_2$) from the training split within a LOSO cross-validation, yielding an interpretable decision rule that partitions the feature value into three regions. We considered four thresholding strategies: \textbf{RuleMedian} and \textbf{RuleAvg}, which compute class representatives (median or mean of $x_f$ within each label) and set $T_1 = (r_0 + r_1)/2, T_2=(r_1 + r_2)/2$; \textbf{RulePercentile}, which sets thresholds using fixed percentiles (33rd/67th) with the direction reversed for COMP to reflect that lower compensation indicates better performance; and \textbf{RuleTree}, which fits a shallow decision tree (max\_depth = 2) from $x_f$ to $y$ and extracts split points to form ($T_1, T_2$). At inference time, each rule outputs both a discrete label and a soft probability vector over $\{0, 1, 2\}$ using element-specific monotonic mappings (i.e. piecewise-linear for ROM and smooth sigmoid transitions for COMP), enabling aggregation across multiple rules. To construct concise per-subject explanations, we ranked candidate rules by their held-out performance and selected the Top-$K$ rules ($K \in \{3, ..., 10\}$) for rule-guided failure detection and hybrid inference/model adaptation. Overall, a compact Top-$3$ rule set learned via \textbf{RuleTree} provided a strong interpretable rule baseline (Appendix. Table \ref{tab:topk_rules_single}) for failure detection and a practical choice for hybrid ML--rule aggregation

For the hybrid of ML and rule-based models, we combined the base ML model prediction with the selected rule outputs using either (i) majority voting over predicted labels or (ii) probability aggregation, which averages the probability vectors from the ML model and rules (uniform weights by default) and predicts $\mathrm{arg}{max}$ of the aggregated distribution. This produces a final label while preserving interpretable rule-level rationales and allowing us to quantify agreement/misalignment between the ML prediction and rule-based evidence. Based on validation results, we use probability aggregation in the user study and fix the rule set to \textbf{RuleTree} with Top-$3$ rules.

\subsubsection{Comparison to Alternative Failure Signals}
To examine whether rule-guided mismatch cues provide value beyond generic AI unreliability cues, we compared our rule-guided disagreement signal against several alternative failure signals computed under the same leave-one-subject-out (LOSO) protocol used throughout the study. Specifically, we evaluated three uncertainty-derived signals from the base neural network output—prediction confidence, prediction margin, and predictive entropy \cite{gawlikowski2023survey}, as well as an ensemble-disagreement baseline \cite{rahaman2021uncertainty} computed from three independently trained model seeds. For each test case, these methods produced a scalar score intended to indicate the likelihood that the AI prediction was incorrect. We then evaluated how well each signal identified AI-error cases by measuring area under the precision–recall curve (AUPRC), treating incorrect AI predictions as the positive class.

\subsubsection{AI Explanations}\label{section:xai}

As the domain experts, therapists prefer feature-oriented explanations for rehabilitation assessment \cite{lee2020co}, the interface includes feature-based explanations to assist case review. For each instance, we computed local feature attributions over our feed-forward NN models using SHAP \cite{NIPS2017Shap} and surfaced only the top three features to manage cognitive load in line with human-AI guidelines \cite{amershi2019guidelines}.

\subsubsection{Interactive Preview Module}
To support low-latency what-if feedback during refinement, we implement an interactive preview module (Figure \ref{fig:interface_phase2}) that estimates how a candidate update would change (i) projected performance and (ii) the embedding space. The preview is computed on demand and \textbf{does not retrain} the underlying neural network model. A preview is triggered whenever a participant proposes an update: label feedback on disagreed cases or rule-based feedback (e.g. selecting rules and adjusting thresholds). 

For projected performance, we use two approaches depending on feedback type. For label feedback, we evaluate a lightweight k-nearest-neighbors classifier on the updated embedding and report local (edited patient) and global (overall) before/after summaries as a fast proxy of change. For rule-based feedback, we apply hybrid inference that combines the frozen ML model with rule outputs (e.g. probability aggregation) to obtain predictions on a held-out evaluation split, and we report the estimated performance change as a pre-commit signal. 

To preview embedding shifts, we re-project NN penultimate-layer embeddings into 2D using supervised Uniform Manifold Approximation and Projection (UMAP) \cite{sainburg2021parametric}, conditioning on the updated labels (from label or rule-based feedback). We then aggregate trial-wise points into patient-side points (median across trials for each patient) and refresh kNN neighborhoods for similar cases inspection. We also summarize class-level movement using per-class centroids in the 2D space. For each class $c \in {0,1,2}$, we compute the centroid as the component-wise median of $(PC1,PC2)$ over participant-side points assigned to that class $c$ to reduce sensitivity to outliers. If a class becomes empty after an edit, we fall back to the last available centroid to avoid unstable behavior.

As the preview is grounded in the neural model’s learned representation, neighborhood and centroid shifts provide an additional check on whether a proposed edit makes cases more locally consistent with nearby examples in embedding space. Together with projected performance change, these views help users inspect the likely consequences of an edit before commitment.

\subsection{Experimental Designs}\label{section:experiments}
In this section, we describe our experimental setup, including study conditions, participants, protocol, and evaluation metrics. Our study materials and procedures were reviewed and approved by the Institutional Review Board (IRB). The study is designed to evaluate two questions: whether rule-guided mismatch cues improve AI-assisted decisions at decision time, and whether prospective preview improves the quality of user-authored edits at refinement time.

\subsubsection{Experimental Conditions}\label{section:experimental_conditions}
We investigate two interface manipulations across two phases: Phase 1 examines rule-based failure detection during AI-assisted decision making and Phase 2 examines feedback-based model adaptation with/without an embedding-shift preview. For Phase 1 (Figure \ref{fig:interface_phase1}), participants completed AI-assisted assessment tasks under two conditions that differed only in whether rule-guided mismatch cues were shown.
\begin{itemize}
    \item Condition A (baseline AI assistance): participants review the AI output (prediction, confidence scores, and explanations) while completing each case. No rule-guided mismatch cues were shown.
    \item Condition B (rule-guided failure inspection): the system includes all baseline elements and additional mismatch cues from a rule table, highlighting potential inconsistencies between the AI prediction and rule-based evidence. These cues were intended to assist participants to identify likely AI failures and calibrate accept/override decisions.
\end{itemize}

For Phase 2, participants provide label and rule-based feedback (e.g. selecting a rule, adjusting rule thresholds) to refine the system (Figure \ref{fig:interface_phase2}). Both conditions showed projected performance change before commitment, but differed in whether the system also visualized embedding shifts.
\begin{itemize}
    \item Condition A (Performance-only preview): after participants propose feedback, the system showed prospective performance changes only to support the commit decision. 
    \item Condition B (Performance + embedding-shift preview): the system showed both projected performance changes and embedding shifts (i.e. how the representation of cases and class centroids would move under the proposed updates) before users decided whether to commit.
\end{itemize}

\subsubsection{Participants}\label{section:participants}
Our final analyzed sample consisted of 21 participants (domain experts and novices), recruited through an advertisement sent to hospital staff, mailing lists, and the research team's professional contacts. Domain experts were rehabilitation therapists with stroke rehabilitation experience (e.g. occupational or physiotherapy practice). Novices were health professionals with limited stroke rehabilitation experience and students majoring in medicine, health or therapy. Our participants includes 9 domain experts (5 occupational therapists and 4 physiotherapists) and 12 novices (1 healthcare worker, 1 therapist, 3 nurses, 3 medical students, 1 nursing student, and 2 physiotherapy students). Detailed demographics are provided in Appendix. Table \ref{tab:participants_details}.

\subsubsection{Protocol}\label{section:protocol}
\begin{figure}[htp]
\centering 
  \centering
  \includegraphics[width=1.0\columnwidth]{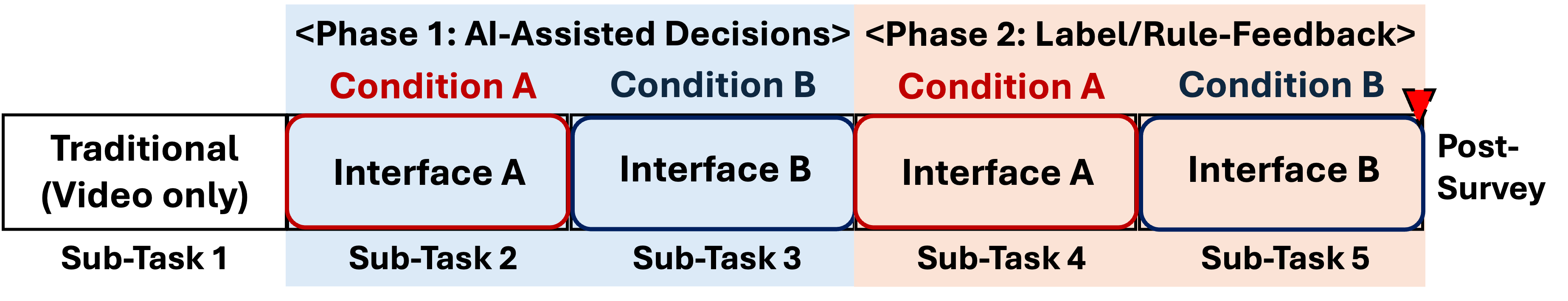}
\caption{Two-phase within-subject study. In Phase 1 (AI-assisted decisions), participants complete tasks under Condition A (baseline) and Condition B with rule-based failure detection. In Phase 2 (Label/rule-based feedback), participants review prospective updates under Condition A (projected performance-only preview) and Condition B (projected performance + embedding shifts preview).}\label{fig:study_flow}
\end{figure}

We conducted a two-phase within-subject design study. Phase 1 examined how rule-guided mismatch cues influenced reliance during AI-assisted decision-making. Phase 2 examined how prospective preview affected the quality of user-authored refinements under performance-only versus performance and embedding-shift previews. After providing informed consent, participants completed a tutorial on the system and procedures and then proceeded through the study using Condition A and Condition B (Figure \ref{fig:study_flow}).

During the tutorial, participants were introduced to the rehabilitation assessment task and labeling criteria, the system implementations, the interface elements (e.g. AI predictions, confidence scores, feature-based explanations, and rule-based interactions), and the overall workflow. After the tutorial, participants first reviewed 14 post-stroke survivors' videos and provided an initial assessment without seeing AI outputs to establish a human-only baseline.

\textbf{Phase 1 (AI-assisted decisions)}: Participants reviewed the AI outputs (predictions, confidence scores, and feature-based explanations). In Condition B, the system additionally displayed rule-guided mismatch cues for failure inspection. After reviewing the AI outputs (and rule-based failure detection), participants submitted a final assessment by either accepting or overriding the AI suggestion (Appendix. Figure \ref{fig:interface_phase1_appendix}). Each participant completed 14 decisions per condition in Phase 1. To ensure that participants encountered both correct and incorrect AI recommendations, we curated cases such that each condition included 10 AI-correct and 4 AI-incorrect outputs (approximately matching the AI model's performance). We counterbalanced the order of Condition A and B across participants and randomized case presentation order within each condition. The AI prediction, confidence, and explanation remained fixed across conditions; the only additional interface element in Condition B was the mismatch cue.

\textbf{Phase 2 (Label/rule feedback for model adaptation)}: Participants provided label and rule-based feedback (e.g. selecting a rule and adjusting thresholds) to improve and adapt the system. For Condition A (performance-only preview), after participants proposed feedback, the system displayed projected performance changes only. For Condition B, (performance + embedding-shift preview), the system additionally visualized embedding shifts. The order of Conditions A and B in Phase 2 was counterbalanced across participants. Phase 2 always followed Phase 1 to preserve a natural workflow from AI-assisted decision-making, decision-time failure detection to refinement. For the cases, in which participants disagreed with the AI recommendation, participants provided label feedback. For rule-based feedback, participants provided six feedback actions across three cases per condition (selected from cases they disagreed with and/or cases with incorrect AI outputs). Finally, participants completed post-study questionnaires and received fixed compensation based on the rate recommended by the domain experts from a national professional organization.

\subsubsection{Metrics for Phase 1}\label{section:metrics}
We build on prior work on human-AI decision-making and reliance \cite{cai2019human,lai2021towards,bansal2021does,lee2025towards,guo2024decision} and utilize performance, reliance, and safety-oriented metrics \cite{lee2026accuracy}. Full formal definitions are provided in Appendix (\ref{appendix:metrics_phase1}). We summarize the metrics used in the main analysis:

For performance, we report the following metrics:
(1) \textbf{Human} (${h_0}$) indicates the percentage of cases where the participant's initial decision matches the ground truth. (2) \textbf{Human--AI Team} ($\mathrm{Acc_{team}}$): indicates the percentage of cases where the participant's final decision after seeing AI matches the ground truth. (3) \textbf{Team gain vs.\ Human} ($\mathrm{TeamGain}$): the improvement from human-only to AI-assisted decision making.
To quantify avoidable errors relative to what was achievable on each trial, we also compute
(4) \textbf{Regret vs.\ best-of} (Regret\_best): the percentage of cases where the final team decision is incorrect despite either the initial human decision or the AI prediction being correct \cite{guo2024decision}. 

To measure reliance quality, we compute accept/reject rates conditioned on AI correctness that measure appropriate reliance (accepting correct AI and reject incorrect AI) and failures, such as over-reliance:
(5) \textbf{Accept-on-correct}: how often participants accepted the AI when the AI was correct, (6) \textbf{Reject-on-wrong}: how often participants overrode the AI when the AI was wrong, (7) \textbf{Reject-on-correct}: how often participants overrode the AI when the AI was correct, and 
(8) \textbf{Accept-on-wrong}: how often participants accepted the AI when the AI was wrong.
As avoiding harm from incorrect reliance is central to safety and responsible deployment, we additionally report (9) \textbf{AI-harm rate}: the percentage of cases where the participant was initially correct but became incorrect after seeing AI. 

To characterize how participants changed their decisions after seeing AI, we also report: (10) \textbf{Changed\_rate}: how often the participant changed their decision after seeing AI, (11) \textbf{ChangedToRight\_rate}: the percentage of cases in which a participant changed an initially wrong decision to a correct one, and 
(12) \textbf{ChangedToWrong\_rate}: the percentage of cases in which a participant changed an initially correct decision to a wrong one. Together, these metrics capture overall outcome improvement, calibrated reliance, and safety-relevant decision changes.

\subsubsection{Phase 2: Interactive Refinement and  Adaptation}
We evaluate how participant feedback improves the system under two preview conditions: 
performance-only preview (Condition A) and performance + embedding-shift preview (Condition B). In Phase~2, participants provide interactive refinement feedback (label feedback on disagreed cases and rule-based feedback by selecting/editing rules and thresholds). 

For re-label feedback, we treat each correction as an additional supervised instance and perform offline adaptation by augmenting the baseline training data with these feedback instances, followed by fine-tuning from the same baseline checkpoint. We apply lightweight quality control (dropping incomplete feedback, deduplicating identical instances, and consolidating conflicting labels with a majority vote).

Rule-based feedback allows participants to select or edit interpretable feature--threshold rules that deterministically adjust the model output at inference time. We implement these edits as a \emph{patient-specific rule layer} composed with the baseline ML predictor without changing the learned model parameters. Rules are authored for a given patient by default; to assess whether they capture broader patterns or are overly specific, we evaluate both within-patient effects and cross-patient transfer by applying the same edited rules to held-out data from other patients.

Across both feedback types, we construct \emph{local} updates using only feedback from a specific patient $i$ (within a condition $c\in\{A,B\}$) and \emph{global} updates using pooled feedback across patients. For re-label feedback, the global update corresponds to a single condition-level fine-tuned model. For rule-based feedback, the global analysis evaluates cross-patient transfer of patient-authored rules. We conduct offline adaptation evaluations to quantify how these feedback mechanisms translate into measurable performance change relative to the same baseline model on a held-out evaluation set.


\section{User Study Results}
We report the results for the two phases of the study. In Phase 1, we evaluate whether rule-guided failure inspection improves AI-assisted decision making. In Phase 2, we examine whether previewing prospective performance changes and embedding shifts helps participants provide higher-quality label and rule-based feedback when refining the model. 
For each metric in Phase 1 and performance improvement measure in Phase 2, we report descriptive statistics and conducted significance tests of  within-participant differences. After assessing data normality using the Kolmogorov-Smirnov test \cite{massey1951kolmogorov}, we conducted paired t-tests for normally distributed metrics and the Wilcox signed-rank tests \cite{cuzick1985wilcoxon} otherwise. 

\subsection{Phase 1: AI-Assisted Decision-Making \& Failure Detection}
\textbf{Comparison to Alternative Failure Signals.} Before examining participants’ decisions, we first compared whether rule-guided mismatch cues identified likely AI failures more effectively than alternative failure signals. Under the same LOSO protocol, uncertainty-based signals from the base model achieved AUPRC values in the range of 0.46–0.49, while ensemble disagreement across three model seeds achieved an AUPRC of 0.52. In contrast, our rule-guided disagreement signal achieved an AUPRC of 0.64. This suggests that rule-derived mismatch captures complementary information not fully reflected in confidence, margin, entropy, or seed-level prediction instability, and provides a plausible mechanism for the improved decision outcomes observed in Phase 1.

We first examine whether exposing mismatch cues improved participants' final decisions after viewing AI. Under Condition~A (baseline AI assistance without rule-guided failure detection), Human--AI Team accuracy increased only marginally from 67.69\% to 69.71\% ($\Delta$=+2.02\%, $p=0.47$ - Appendix. Table \ref{tab:performance_all_detail}). In contrast, under Condition~B (rule-guided failure detection), Human--AI Team accuracy increased significantly from 67.86\% to 82.02\% ($\Delta$=+14.16\%, $p<0.001$ - Appendix. Table \ref{tab:performance_all_detail}). 

Comparing conditions directly (Table~\ref{tab:results_perf_teamgains_transposed}), Condition~B (rule-guided failure detection) yielded significantly higher Human--AI Team accuracy than Condition~A ($\Delta$=+12.31\%, $p<0.001$) and significantly larger team gain vs.\ human-only decisions ($\Delta$=+12.14\%, $p<0.01$). Rule-guided failure detection (Condition~B) also reduced avoidable errors as measured by Regret\_best (16.46\% $\rightarrow$ 4.80\%; $\Delta$=-11.66\%, $p<0.01$), indicating fewer cases where the final team decision was incorrect even though either the initial human decision or the AI was correct.

\begin{table*}[htp]
\centering
\caption{Performance and team-gain metrics. Compared with baseline AI assistance (Condition A), rule-guided mismatch cues (Condition B) significantly improved Human-AI Team accuracy and team gain, while reducing avoidable errors, Regret\_best.}
\label{tab:results_perf_teamgains_transposed}
\resizebox{0.62\textwidth}{!}{%
\begin{tabular}{lcccc}
\toprule
\textbf{} &
\textbf{Human} &
\textbf{Human--AI Team} &
\textbf{\begin{tabular}[c]{@{}c@{}}Team gain vs. Human\end{tabular}} &
\textbf{Regret\_best} \\
\midrule
\textbf{Condition A} & 67.69 & 69.71 & 2.02  & 16.46 \\ \midrule
\textbf{Condition B} & 67.86 & 82.02 & 14.16 & 4.80  \\ \midrule
\textbf{\begin{tabular}[c]{@{}l@{}}$\Delta$ B-A (p-value)\end{tabular}} &
\begin{tabular}[c]{@{}c@{}}0.17 (p=0.86)\end{tabular} &
\begin{tabular}[c]{@{}c@{}}$\uparrow$ 12.31 (p$<$0.001)\end{tabular} &
\begin{tabular}[c]{@{}c@{}}$\uparrow$ 12.14 (p$<$0.01)\end{tabular} &
\begin{tabular}[c]{@{}c@{}}$\downarrow$ -11.66 (p$<$0.01)\end{tabular} \\
\bottomrule
\end{tabular}%
}
\end{table*}

\begin{table*}[htp]
\centering
\caption{Reliance metrics conditioned on AI correctness. Rule-guided mismatch cues improved  reliance calibration by increasing acceptance of correct AI and rejection of incorrect AI, while reducing both over/under-reliance and AI-induced harm.}
\label{tab:results_reliance_transposed}
\resizebox{0.7\textwidth}{!}{%
\begin{tabular}{lccccc}
\toprule
\textbf{} &
\textbf{\begin{tabular}[c]{@{}c@{}}Accept\\ on\_correct\end{tabular}} &
\textbf{\begin{tabular}[c]{@{}c@{}}Reject\\ on\_wrong\end{tabular}} &
\textbf{\begin{tabular}[c]{@{}c@{}}Reject\\ on\_correct\end{tabular}} &
\textbf{\begin{tabular}[c]{@{}c@{}}Accept\\ on\_wrong\end{tabular}} &
\textbf{\begin{tabular}[c]{@{}c@{}}AI\_harm\\ rate\end{tabular}} \\
\midrule
\textbf{ConditionA} & 88.09 & 41.61 & 11.91 & 58.39 & 8.29 \\ \midrule
\textbf{ConditionB} & 95.24 & 62.14 &  4.76 & 37.86 & 3.70 \\ \midrule
\textbf{\begin{tabular}[c]{@{}l@{}}$\Delta$ B-A (p-value)\end{tabular}} &
\begin{tabular}[c]{@{}c@{}}$\uparrow$ 7.15  (p$<$0.01)\end{tabular} &
\begin{tabular}[c]{@{}c@{}}$\uparrow$ 20.53  (p$<$0.05)\end{tabular} &
\begin{tabular}[c]{@{}c@{}}$\downarrow$ -7.15 (p$<$0.01)\end{tabular} &
\begin{tabular}[c]{@{}c@{}}$\downarrow$ -20.53 (p$<$0.01)\end{tabular} &
\begin{tabular}[c]{@{}c@{}}$\downarrow$ -4.59  (p$<$0.05)\end{tabular} \\
\bottomrule
\end{tabular}%
}
\end{table*}

\textbf{Subgroup trends} (Appendix. Tables \ref{tab:performance_all_detail} and  \ref{tab:results_teamgains_detail}).
The same pattern held for both experts and novices. Experts showed no improvement of Human-AI Team under Condition~A (71.35\% $\rightarrow$ 70.19\%, $p=0.56$), but improved significantly in Condition~B (72.22\% $\rightarrow$ 83.43\%, $p<0.05$). Novices improved modestly the Human-AI Team metric under Condition~A (64.94\% $\rightarrow$ 69.35\%, $p=0.18$; n.s.) and improved significantly under Condition~B (64.58\% $\rightarrow$ 80.95\%, $p<0.01$). Across subgroups, Regret\_best decreased in Condition~B (Appendix Table~\ref{tab:results_teamgains_detail}), suggesting that mismatch cues assisted participants to better select between their initial decision and the AI recommendation, thereby reducing avoidable errors even though oracle-best accuracy (i.e. the availability of a correct answer from either human or AI) remained similar across conditions. 

\subsubsection{Reliance Behavior (Accept/Reject Patterns)}

To understand how mismatch cues changed reliance behavior, we analyze accept/reject behaviors conditioned on AI correctness (Table~\ref{tab:results_reliance_transposed}). Compared to the baseline (Condition~A), rule-guided failure detection (Condition~B) significantly increased \textit{Accept-on-correct} (88.09\% $\rightarrow$ 95.24\%, $\Delta$=+7.15\%, $p<0.01$) and \textit{Reject-on-wrong} (41.61\% $\rightarrow$ 62.14\%, $\Delta$=+20.53\%, $p<0.05$). In parallel, it significantly decreased \textit{Reject-on-correct} (11.91\% $\rightarrow$ 4.76\%, $\Delta$=-7.15\%, $p<0.01$) and \textit{Accept-on-wrong} (58.39\% $\rightarrow$ 37.86\%, $\Delta$=-20.53\%, $p<0.01$). Taken together, these results indicate better reliance calibration under Condition B (rule-guided failure detection): participants were more likely to follow correct AI, \textbf{more likely to reject incorrect AI}, and less likely to show either over-reliance or under-reliance. 

As a safety-oriented reliance measure, rule-guided failure detection (Condition~B) also reduced the \textit{AI-harm rate} (8.29\% $\rightarrow$ 3.70\%, $\Delta$=-4.59\%, $p<0.05$), indicating fewer cases where participants moved from an initially correct decision to an incorrect final decision after viewing AI.

\textbf{Subgroup trends.} (Appendix Table~\ref{tab:results_reliance_detail}) Experts showed strong improvements in \textit{Accept-on-correct} (86.40\% $\rightarrow$ 98.89\%, $p<0.01$) and reductions in \textit{Reject-on-correct} (13.60\% $\rightarrow$ 1.11\%, $p<0.01$). For novices, the accept/reject metrics changed in the same direction, such as higher \textit{Reject-on-wrong} and lower \textit{Accept-on-wrong}, but these differences were not statistically significant (often $p\approx 0.1$), consistent with lower power in subgroup analyses. Help-oriented metrics (e.g. \textit{AI-help}, \textit{missed-help}, \textit{correct-ignore}) were directionally positive under rule-guided failure detection (Condition B) (i.e. slightly higher AI-help and lower missed-help) but were not statistically significant.

\subsubsection{Decision-Change Behavior After Viewing AI}

We then exmaine how participants changed their decisions after seeing AI, in order to better understand how the improvements above were achieved (Table~\ref{tab:results_changed_decisions_transposed}). Overall, rule-guided failure detection (Condition~B) reduced the \textit{Changed} rate (34.98\% $\rightarrow$ 26.66\%, $\Delta$=-8.32\%, $p<0.1$), while showing a directional increase in \textit{ChangedToRight} (14.74\% $\rightarrow$ 18.88\%, $\Delta$=+4.14\%, $p=0.13$). In addition, rule-guided failure detection (Condition~B) significantly reduced \textit{ChangedToWrong} decisions (12.72\% $\rightarrow$ 4.72\%, $\Delta$=-8.00\%, $p<0.01$), indicating that rule-guided failure detection reduced harmful switches from initially correct judgments to incorrect final decisions.

\begin{table}[htp]
\centering
\caption{Decision-change behaviors after viewing AI outputs. Rule-guided mismatch cues (Condition B) reduced overall switching and significantly reduced harmful changes from correct initial judgments to incorrect final decisions.}
\label{tab:results_changed_decisions_transposed}
\resizebox{0.6\linewidth}{!}{%
\begin{tabular}{lccc}
\toprule
\textbf{} &
\textbf{\begin{tabular}[c]{@{}c@{}}Changed\end{tabular}} &
\textbf{\begin{tabular}[c]{@{}c@{}}ChangedToRight\end{tabular}} &
\textbf{\begin{tabular}[c]{@{}c@{}}ChangedToWrong \end{tabular}} \\
\midrule
\textbf{ConditionA} & 34.98 & 14.74 & 12.72 \\ \midrule
\textbf{ConditionB} & 26.66 & 18.88 &  4.72 \\ \midrule
\textbf{\begin{tabular}[c]{@{}l@{}}$\Delta$ B-A  (p-value)\end{tabular}} &
\begin{tabular}[c]{@{}c@{}}$\downarrow$-8.32  (p$<$0.1)\end{tabular} &
\begin{tabular}[c]{@{}c@{}}4.14 (p=0.13)\end{tabular} &
\begin{tabular}[c]{@{}c@{}}$\downarrow$-8.00  (p$<$0.01)\end{tabular} \\
\bottomrule
\end{tabular}%
}
\end{table}

\textbf{Subgroup trends.}
(Appendix Table~\ref{tab:results_changed_decisions_detail}) The reduction in \textit{ChangedToWrong} holds for both experts (13.73\% $\rightarrow$ 6.25\%, $p<0.05$) and novices (11.96\% $\rightarrow$ 3.57\%, $p<0.05$). While the \textit{Changed} and \textit{ChangedToRight} showed similar directional patterns (increases) within subgroups, these effects were not statistically significant, again consistent with reduced power.

\subsection{Phase 2: Model Adaptation}
We analyzed how participants' feedback affected model refinement under the two preview conditions
(Table~\ref{tab:phase2_refinement_all}). 
Overall, Phase~2 refinement results show a clear asymmetry between \emph{local} versus \emph{global} updating (Table~\ref{tab:phase2_refinement_all}). Local (participant-specific) updates improved performance relative to the baseline in both conditions, with significantly larger gains in Performance+Embedding-Shift Preview (Condition~B) (A: \textbf{+11.50} vs B: \textbf{+36.38}; $\Delta$B--A=\textbf{+24.89},  $p<0.001$). In contrast, global (across patients) updates did not provide reliable improvement: Performance-only Preview (Condition~A) produced only a small gain (\textbf{+2.38}), while Performance+Embedding-Shift Preview (Condition~B) performed below baseline (\textbf{-5.05}), yielding a significant negative difference ($\Delta$B--A=\textbf{-7.43}, $p<0.001$). Together, these findings suggest that richer preview is particularly useful for local personalization while revealing that edits beneficial at the local level may not remain beneficial when aggregated into a single global update. 

\textbf{Subgroup trends.} (
Appendix Table~\ref{tab:phase2_refinement}) The local-update advantage of Performance+Embedding-Shift Preview (Condition~B) held for both experts (A: \textbf{+11.20} $\rightarrow$ B: \textbf{+33.89}, $\Delta=\textbf{+22.69}$, $p<0.001$) and novices (A: \textbf{+11.92} $\rightarrow$ B: \textbf{+38.88}, $\Delta=\textbf{+26.96}$, $p<0.05$). For global updates, both subgroups show the same directional pattern of degradation under Condition~B relative to Condition~A (experts: \textbf{+5.58} $\rightarrow$ \textbf{-2.95}, $p<0.001$; novices: \textbf{0.00} $\rightarrow$ \textbf{-6.46}, $p<0.01$), consistent with pooled updates being less stable than personalized refinement.

\begin{table}[t]
\centering
\caption{Phase 2 refinement performance for all participants. We report improvement over the baseline model on a held-out evaluation set for local (participant-specific) and global (across-patients) updates. $p$-values are from paired Wilcoxon signed-rank tests across participants.}
\label{tab:phase2_refinement_all}
\small
\begin{tabular}{lcccc}
\toprule
Metric & A & B & $\Delta$B--A & $p$ \\
\midrule
Local update $\Delta$Perf (overall)  & 11.50 & 36.38 & 24.89 & $<0.001$ \\
Global update $\Delta$Perf (overall) & 2.38  & -5.05 & -7.43 & $<0.001$ \\
\bottomrule
\end{tabular}
\end{table}

Additional analysis of global direct retraining showed that ROM improved only slightly (best $\Delta\approx +1.99$), while COMP degraded substantially (best $\Delta\approx -14.92$), leading to an overall decrease in performance (best $\Delta\approx -6.46$). This pattern indicates that naively aggregating feedback into a single global update can introduce task-specific trade-offs and may require additional safeguards (e.g. per-task updates, data curation, or constraints) to avoid regressions. In contrast, {rule-based refinement yields strong \emph{local} gains but limited cross-patient transfer.}

\section{Discussion}
Our findings suggest that failure-aware human-AI model editing benefits from two complementary interface capabilities: decision-time mismatch cues and refinement-time prospective preview. In Phase 1, rule-guided mismatch cues improved Human–AI team performance and reliance calibration by helping participants decide when to accept or override the AI. In Phase 2, prospective preview improved the quality of local refinements, while also revealing that edits that appear beneficial locally may not remain beneficial when pooled into broader global updates. Together, these results shift the design focus from merely showing AI outputs to supporting failure inspection, user-authored intervention, and pre-commit reasoning about model change, enabling  more actionable and anticipatory human-AI collaboration.

\subsection{Rule-Guided Mismatch Cues Support Calibrated Reliance}
Our Phase~1 findings show that rule-guided mismatch cues can calibrate reliance by helping users decide \emph{when to follow} AI and \emph{when to override} it. Adding interpretable mismatch cues from rule-based evidence (Condition~B) yielded a large improvement in Human--AI Team accuracy compared to baseline AI assistance (Condition~A): 82.02\% vs.\ 69.71\% ($\Delta$=+12.31\%, $p<0.001$), while human-only performance did not differ across conditions (67.86\% vs.\ 67.69\%, $p=0.86$). 
This extends prior work showing that AI outputs (e.g. explanations and confidence scores) can induce overreliance: beyond helping users interpret the AI, a lightweight mismatch cue can help them act on potential failures at decision time. 

Beyond team performance, rule-guided mismatch cues also delivered measurable safety benefits. They significantly reduced harmful reliance events, lowering AI-induced harm (\textit{AI-harm}: 8.29\% $\rightarrow$ 3.70\%, $p<0.05$) and reduced harmful switches from correct initial judgments to incorrect final decisions (\textit{ChangedToWrong}: 12.72\% $\rightarrow$ 4.72\%, $p<0.01$). In addition, rule-guided mismatch cues reduced avoidable errors in a decision-theoretic sense: Regret\_best dropped from 16.46\% to 4.80\% ($p<0.01$), meaning teams were less likely to end wrong even when either the human or AI had the correct answer available \cite{guo2024decision}. Together, these results suggest that \emph{interpretable consistency checks} can act as a practical guardrail for both reliance calibration and harm reduction, complementing prior approaches that emphasize richer explanations or uncertainty displays \cite{bansal2021does,lee2025towards,li2025confidence,kim2025fostering}.

Rules functioned as simple, interpretable heuristics that prompted participants to scrutinize potentially inconsistent AI outputs, supporting reflective rather than automatic acceptance. This resonates with recent findings on confidence alignment \cite{li2025confidence,chen2023understanding} and uncertainty-aware delegation \cite{lee2025towards}, yet extends them by offering a structured and interpretable mechanism for trust calibration. A key next step is to understand how such cues should be learned or specified, validated, and personalized across different error profiles, tasks, and domains. Another is to examine how mismatch-based signals can be combined with uncertainty-aware or decision theoretic reliance objectives to jointly optimize accuracy and safety \cite{guo2024decision}.

\subsection{Prospective Previews for Anticipatory Model Refinement}
Our Phase~2 results reveal that presenting prospective embedding updates during model refinement improved participants' ability to provide targeted, high-quality feedback. When participants could inspect how their selected rules would reshape the representation space (e.g. shifts in class centroids and data points) alongside projected performance changes, the resulting \emph{local} (per-patient) improvement increased substantially (Condition~A: 11.50\% $\rightarrow$ Condition~B: 36.38\%, $p<0.001$). This suggests that preview helps users reason about the likely consequences of an edit before commitment, enabling more targeted refinements that are better aligned with the model's learned representation. At the same time, these benefits were primarily \emph{local}: when feedback was aggregated for a single \emph{global} update, performance did not reliably improve and could regress (global $\Delta$Perf: Condition~A \textbf{+2.38} vs Condition~B \textbf{-5.05}, $p<0.001$; Table~\ref{tab:phase2_refinement_all}). Consistent with our global direct-retraining analysis, pooled global updates can introduce task-specific trade-offs (e.g. COMP requires patient-specific assessment \cite{lee2020co}), yielding a net decrease in performance. This local-global asymmetry indicates that systems supporting user-authored editing should not assume that locally beneficial edits are globally safe. Instead, broader deployment may require safeguards \cite{bengio2025international}, such as per-task updates, stricter data curation, conflict-aware aggregation, or constrained optimization (e.g. partial freezing, conservative learning rates). 

We interpret this effect as a proactive form of anticipatory calibration and alignment: users learn not only from retrospective outcomes (i.e. observing an error after it occurs), but also from prospective consequences (i.e. previewing how an intervention will change the model) before an intervention is committed. Prior interactive and explanatory debugging systems largely emphasize retrospective error correction (i.e. users identify failures, provide corrective feedback, and then observe the updated behavior) \cite{kulesza2015principles,lee2021human,guo2022building}. In contrast, our approach makes the likely {direction} and {magnitude} of change legible before commitment, supporting planning-oriented interaction and helping users reason about potential failure propagation in advance. By improving users' mental models of how feedback translates into model change, prospective preview can support more transparent and controllable refinement for trustworthy human-AI collaboration, while reducing reliance on trial-and-error editing \cite{holstein2023toward,buccinca2021trust}.

\subsection{From Explanations to Failure-Aware Model Editing}
Our study highlights a shift from explanation-centered systems towards failure-aware model editing. Traditional explainable AI (XAI) methods, such as feature importance \cite{lee2021human,wang2021explanations}, counterfactual \cite{lee2023understanding,zhang2022towards} or uncertainty scores \cite{zhang2020effect,prabhudesai2023understanding,li2025confidence,lee2025towards}, offer informational transparency \cite{wang2019designing,arya2019one}, but they often fail to foster calibrated reliance or provide users with mechanisms to intervene when the system fails. RuleEdit connects explanation-like evidence to controllable mechanisms, such as rejecting a likely-wrong AI output or proposing targeted feedback for refinement. Through these mechanisms, users move from passively interpreting AI outputs to actively inspecting likely failures and editing model behavior.

This approach aligns with interactive auditing \cite{tang2024ai} and human-centered model governance \cite{green2019principles} by enabling users to trace, inspect, and justify their interventions. Importantly, our controllable interaction moves beyond documenting failures \cite{tang2024ai} to mitigating them: instead of post-hoc interpretation, participants can intervene on failures in situ and inspect prospective consequences before commitment.

\subsection{Limitations and Future Work}
While our findings provide promising evidence for rule-guided failure detection and model refinement, this study also has several limitations. First, our study was conducted in a single domain, stroke rehabilitation assessment with a modest sample size. Although this scale is comparable to prior work on human-AI collaborative decision-making in health \cite{lee2020co,bussone2015role}, future work should examine whether observed improvements in reliance calibration and model refinement generalize to other domains, such as diagnostic imaging \cite{rajpurkar2022ai} or child welfare services \cite{de2020case,stapleton2022imagining}. 

In addition, the model adaptation process was evaluated offline rather than within real or longitudinal decision-making workflows. Field deployment studies would provide deeper insights into how trust, learning, and editing behavior evolve over time. As rehabilitation practices are typically led by individual therapists, our study focused on individual feedback. Future work should extend this setting to multi-user workflows in which teams co-create, review, and validate rules and edits for broader organizational use and accountability \cite{novelli2024accountability}.

\section{Conclusion}
In this work, we examined how failure-aware human-AI model editing can be supported through two complementary capabilities: rule-guided mismatch cues and prospective impact preview. Our findings show that rule-guided mismatch cues improved AI-assisted decision-making by helping users identify likely AI failures, decide when to accept or override the AI, and reduce both overreliance and underreliance. Furthermore, prospective preview substantially improved the quality of local user-authored refinements, demonstrating that users' structured, rule-based feedback can meaningfully enhance AI systems in a controlled and auditable workflow.

At the same time, our results reveal an important boundary: edits that appear beneficial locally may not remain beneficial when pooled into broader global updates. This local-global tradeoff underscores the need for supporting pre-commit reasoning about model change rather than trial-and-error editing alone. 

Overall, RuleEdit points toward a shift from explanation-centered AI support to failure inspection, user-authored intervention, and prospective reasoning about model behavior. By linking interpretable rules to controllable editing workflows, our work contributes to the design of more transparent, controllable, and trustworthy human-AI systems.

\bibliographystyle{ACM-Reference-Format}
\bibliography{main}

\appendix
\section{System Implementation Details}
\subsection{Feature Extractions}\label{appendix:feature_extract}
We extend prior rehabilitation assessment work \cite{lee2019learning} by transforming tracked 3D joint trajectories of post-stroke survivors' exercises into kinematic features.

For Range of Motion (ROM), we derive joint-angle features (e.g., elbow flexion/extension, shoulder flexion) and distance related features. Specifically, we compute pairwise Euclidean distances between key joints, such as head–wrist and head–elbow, and wrist distances to the head and shoulder, along with their per axis components ($x,y,z$) \cite{lee2019learning}.
For Compensation, we quantify non-target movement by measuring normalized displacements of the head, spine, and shoulder from the trial’s initial frame over time. These per-axis ($x,y,z$) trajectories summarize behaviors, such as shoulder elevation or trunk lean that indicate compensatory strategies \cite{lee2019learning}.

\subsection{Results of Machine Learning Models}\label{appendix:results_ml}
We trained the feed-forward neural network (NN) model through a grid search over depth (1-4 hidden layers with 32, 64, 128, 256, 512 hidden units) along with learning rates (i.e. 1e-5, 5e-4, 1e-4, 5e-3, 1e-3, 1e-2). Models were trained by minimizing cross-entropy loss and evaluated with leave-one-subject-out cross-validation: in each fold, all data from a single participant were held out for testing, and the model was trained on the remaining data. 

\subsection{Results of Rule-based Models}\label{appendix:results_rb}
Across all three experiments on ROM, COMP, and Overall, performance of rule-based models improves monotonically as we reduce $K$ (Top 10 $\rightarrow$ Top 3), indicating that a small set of high-quality rules is more effective than including weaker rules. For the ROM performance component, RuleMedian achieves the highest performance of 0.9574 at Top3 (vs. RuleTree 0.9297), while for the COMP performance component, RuleTree achieves the best Top3 score (0.8961). Overall, the best Top3 performance is 0.9129 (RuleTree), suggesting that selecting a compact Top3 rule set can yield a strong, interpretable rule baseline and a practical candidate for hybrid of ML and rules aggregation.

\begin{table*}[t]
\centering
\caption{Top-$K$ rule performance across rule construction methods for ROM, COMP, and Overall.}
\label{tab:topk_rules_single}
\small
\setlength{\tabcolsep}{3.5pt}

\resizebox{\textwidth}{!}{%
\begin{tabular}{lcccccccccccc}
\toprule
& \multicolumn{4}{c}{\textbf{ROM}} & \multicolumn{4}{c}{\textbf{COMP}} & \multicolumn{4}{c}{\textbf{Overall}} \\
\cmidrule(lr){2-5}\cmidrule(lr){6-9}\cmidrule(lr){10-13}
Top-$K$ &
RuleMedian & RuleAvg & RulePercentile & RuleTree &
RuleMedian & RuleAvg & RulePercentile & RuleTree &
RuleMedian & RuleAvg & RulePercentile & RuleTree \\
\midrule
Top10 & 0.8544 & 0.8548 & 0.6505 & 0.8094 & 0.8455 & 0.8503 & 0.8311 & 0.8226 & 0.8500 & 0.8525 & 0.7408 & 0.8160 \\
Top9  & 0.8701 & 0.8673 & 0.6719 & 0.8311 & 0.8496 & 0.8533 & 0.8369 & 0.8312 & 0.8598 & 0.8603 & 0.7544 & 0.8312 \\
Top8  & 0.8833 & 0.8774 & 0.6915 & 0.8442 & 0.8522 & 0.8559 & 0.8410 & 0.8407 & 0.8678 & 0.8667 & 0.7663 & 0.8424 \\
Top7  & 0.8970 & 0.8883 & 0.7154 & 0.8599 & 0.8547 & 0.8587 & 0.8445 & 0.8505 & 0.8759 & 0.8735 & 0.7800 & 0.8552 \\
Top6  & 0.9077 & 0.8983 & 0.7439 & 0.8769 & 0.8574 & 0.8612 & 0.8471 & 0.8593 & 0.8826 & 0.8797 & 0.7955 & 0.8681 \\
Top5  & 0.9216 & 0.9113 & 0.7690 & 0.8918 & 0.8594 & 0.8647 & 0.8507 & 0.8686 & 0.8905 & 0.8880 & 0.8098 & 0.8802 \\
Top4  & 0.9378 & 0.9288 & 0.8020 & 0.9074 & 0.8612 & 0.8661 & 0.8539 & 0.8809 & 0.8995 & 0.8975 & 0.8279 & 0.8942 \\
Top3  & \textbf{0.9574} & \textbf{0.9476} & \textbf{0.8400} & \textbf{0.9297} & \textbf{0.8630} & \textbf{0.8668} & \textbf{0.8574} & \textbf{0.8961} & \textbf{0.9102} & \textbf{0.9072} & \textbf{0.8487} & \textbf{0.9129*} \\
\bottomrule
\end{tabular}%
}
\end{table*}

\subsection{Detail of Data Analysis Metrics for Phase 1}\label{appendix:metrics_phase1}
Let $y$ denote the ground-truth label, $h_0$ the participant’s initial (human-only) decision, $a$ the AI prediction, and $h_1$ the participant’s final decision after viewing the AI output (and optionally overriding). We use the indicator function $\mathbb{I}[\cdot]$.

\begin{itemize}
    \item \textbf{Human} (${h_0}$): $\mathbb{I}[h_0 = y]$ indicates the percentage of cases where the participant's initial decision matches the ground truth.
    \item \textbf{AI accuracy} ($\mathrm{Acc}_{ai}$): $\mathbb{I}[a=y]$ indicates the percentage of cases where the AI prediction matches the ground truth.
    \item \textbf{Human--AI Team} ($\mathrm{Acc_{team}}$): $\mathbb{I}[h_1 = y]$ indicates the percentage of cases where the participant's final decision after seeing AI matches the ground truth. 
    \item \textbf{Team gain vs.\ Human} ($\mathrm{TeamGain_{human}}$): $\mathrm{Acc_{team}}$ - $\mathrm{h_0}$, the improvement from human-only to AI-assisted decision making.
    \item \textbf{Team gain vs.\ AI} ($\mathrm{TeamGain_{ai}}$): $\mathrm{Acc_{team}}-\mathrm{Acc_{ai}}$, 
    the improvement from AI accuracy to the team’s final accuracy, computed as $\mathrm{Acc_{team}} - \mathrm{Acc_{ai}}$.
    \item \textbf{Regret\_best} ($\mathrm{Oracle} - \mathbb{I}[h_1=y]$): the percentage of cases where the final team decision is incorrect despite either the initial human decision or the AI prediction being correct \cite{guo2024decision}. This equals 1 when the team is wrong despite at least one agent (human-only or AI) being correct, and 0 otherwise \cite{guo2024decision}.

    \item \textbf{Oracle best accuracy} ($\mathrm{Acc}_{oracle}$): $\mathbb{I} [(h_0=y)\lor(a=y)]$, Upper bound accuracy if one could always choose the correct answer between the initial human decision and the AI prediction.
\end{itemize}

In addition, we compute {accept/reject rates conditioned on AI correctness} that measure appropriate reliance (accepting correct AI and reject incorrect AI) and failures, such as over-reliance:
\begin{itemize}
    \item \textbf{Accept-on-correct} ($\Pr(h_1=a \mid a=y)$) indicates when the participant's final decision agrees with AI among cases with correct AI outputs.
    \item \textbf{Reject-on-wrong}: $\Pr(h_1\neq a \mid a\neq y)$ indicates the percentage where the participant's final decision disagree with AI among cases with wrong AI outputs.
    \item \textbf{Reject-on-correct}: $\Pr(h_1\neq a \mid a=y)$ indicates the percentage where the participant's final decision disagrees with AI among cases with correct AI outputs.
    \item \textbf{Accept-on-wrong}: $\Pr(h_1=a \mid a\neq y)$ indicates the percentage where the participant's final decision agrees with AI among cases with wrong AI outputs
\end{itemize}

To further characterize safety-relevant reliance outcomes, we report a help--harm decomposition. $N$ the number of cases for a participant in a given condition.
\begin{itemize}
    \item \textbf{AI-help rate} ($\frac{1}{N}\sum_{j=1}^{N}\mathbb{I}\big[h_{0j}\neq y_j \ \land\ h_{1j}=y_j\big]$): quantifies cases where the participant is initially incorrect but becomes correct after reviewing AI
    \item \textbf{AI-harm rate} ($\frac{1}{N}\sum_{j=1}^{N}\mathbb{I}\big[h_{0j}=y_j \ \land\ h_{1j}\neq y_j\big]$): quantifies cases where the participant is initially correct but becomes incorrect after reviewing AI.
    \item \textbf{Missed-help rate} ($\frac{1}{N}\sum_{j=1}^{N}\mathbb{I}\big[h_{0j}\neq y_j \ \land\ a_j=y_j \ \land\ h_{1j}\neq a_j\big]$): captures cases where the participant is initially incorrect and the AI prediction is correct, but the participant does not follow the AI and remains incorrect.
    \item \textbf{Correct-ignore rate} ($\frac{1}{N}\sum_{j=1}^{N}\mathbb{I}\big[h_{0j}=y_j \ \land\ a_j\neq y_j \ \land\ h_{1j}\neq a_j\big]$): captures cases where the participant is initially correct and the AI prediction is incorrect, and the participant rejects the AI and remains correct.
\end{itemize}

To characterize when improvements arise and whether decision changes introduce harm, we report decision-change behaviors (Changed/ChangedToRight/ChangedToWrong): 
\begin{itemize}
    \item \textbf{Changed\_rate} ($\mathbb{I}[h_1 \neq h_0]$) indicates the percentage of cases where the participant changes their initial decision after viewing AI. 
    \item \textbf{ChangedToRight\_rate}  $($ $\mathbb{I}[h_1 \neq h_0] \cdot \mathbb{I}[h_0 \neq y] \cdot \mathbb{I}[h_1 = y]$) indicates the percentage of cases where the participant changes their decision and ends up correct.
    \item \textbf{ChangedToWrong\_rate} ($\mathbb{I}[h_1 \neq h_0] \cdot \mathbb{I}[h_0 = y] \cdot \mathbb{I}[h_1 \neq y]$) indicates the percentage of cases where the participant changes their decision and ends up incorrect.
\end{itemize}

\begin{table*}[htp]
\centering
\caption{Detailed Demographics of Participants: Domain experts, therapists, who have experience in stroke rehabilitation more than one year (TP1 - TP9) and novices (NV1 - NV12) (i.e. other health professionals and students majoring in medicine, therapy)}
\label{tab:participants_details}
\resizebox{0.7\textwidth}{!}{%
\begin{tabular}{cccllc} \\ \toprule
\textbf{ID} &
  \textbf{Age} &
  \textbf{Gender} &
  \multicolumn{1}{c}{\textbf{Role}} &
  \multicolumn{1}{c}{\textbf{Setting / Others}} &
  \textbf{Experience} \\ \midrule
TP1  & 35 - 44 years & Female & Occupational Therapist (OT) & Inpatient Rehabilitation      & 13 \\
TP2  & 25 - 34 years & Female & Occupational Therapist (OT) & Outpatient Clinic             & 8  \\
TP3  & 25 - 34 years & Female & Occupational Therapist (OT) & Outpatient Clinic             & 9  \\
TP4  & 25 - 34 years & Female & Occupational Therapist (OT) & Skilled Nursing Facility      & 6  \\
TP5  & 25 - 34 years & Male   & Occupational Therapist (OT) & Other (please specify)        & 7  \\
TP6  & 55 - 64 years & Female & PhysioTherapist (PT)        & Skilled Nursing Facility      & 30 \\
TP7  & 25 - 34 years & Male   & PhysioTherapist (PT)        & Outpatient Clinic             & 2  \\
TP8  & 35 - 44 years & Female & PhysioTherapist (PT)        & Outpatient   Clinic           & 4  \\
TP9  & 35 - 44 years & Male   & PhysioTherapist (PT)        & Skilled Nursing Facility      & 12 \\
NV1  & 25 - 34 years & Male   & Student-Physiotherapy       &                               &    \\
NV2  & 18 - 24 years & Female & Student-Physiotherapy       &                               &    \\
NV3  & 25 - 34 years & Female & Speech Therapist            & Private Therapy Centre        & 2  \\
NV4  & 35 - 44 years & Female & Nurse                       & Home Care Centre              & 13 \\
NV5 &
  25 - 34 years &
  Male &
  Nurse &
  \begin{tabular}[c]{@{}l@{}}Acute Care Centre\\ (experience with \\ post-stroke survivors)\end{tabular} &
  7 \\
NV6  & 25 - 34 years & Male   & Nurse                       &  \begin{tabular}[c]{@{}l@{}}Acute Care Centre\\ (experience with \\ post-stroke survivors)\end{tabular}                             & 8  \\
NV7  & 25 - 34 years & Female & Nurse                       & Nursing home                  & 10 \\
NV8  & 25 - 34 years & Male   & Healthcare Worker           & Community Hospital, Inpatient & 6  \\
NV9  & 18 - 24 years & Female & Student-Medicine            &                               &    \\
NV10 & 18 - 24 years & Male   & Student-Medicine            &                               &    \\
NV11 & 18 - 24 years & Female & Student-Medicine            &                               &    \\
NV12 & 18 - 24 years & Female & Student-Nursing             &                               &    \\ \bottomrule
\end{tabular}%
}
\end{table*}

\begin{figure*}[htp]
\centering 
\includegraphics[width=0.6\textwidth]%
  {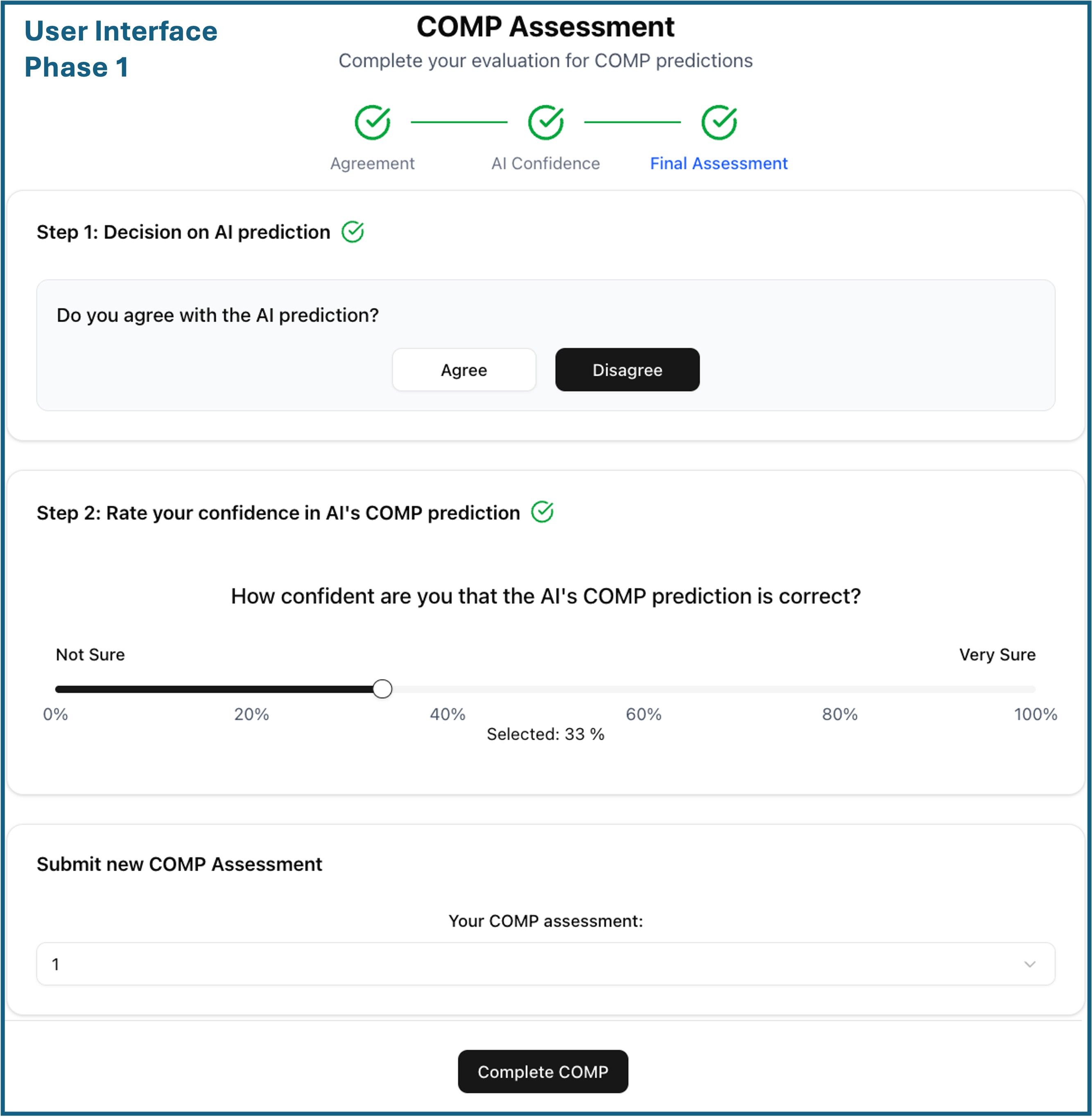}
\caption{User Interface of Phase 1.}\label{fig:interface_phase1_appendix}
\end{figure*}

\begin{table*}[htp]
\centering
\caption{Performance of decision-making without/with AI-assistance by all participants (All), domain experts (EXPs), and health professionals and students (NVs)}
\label{tab:performance_all_detail}
\resizebox{0.9\textwidth}{!}{%
\begin{tabular}{cccccccccc} \toprule
 & \multicolumn{3}{c}{\textbf{All}}                      & \multicolumn{3}{c}{\textbf{Experts}}                  & \multicolumn{3}{c}{\textbf{Novices}}                  \\
 & \textbf{Human} & \textbf{Human-AI} & \textbf{Changes} & \textbf{Human} & \textbf{Human-AI} & \textbf{Changes} & \textbf{Human} & \textbf{Human-AI} & \textbf{Changes} \\ \midrule
\textbf{ConditionA} &
  67.69 &
  69.71 &
  \begin{tabular}[c]{@{}c@{}}$\uparrow$ 2.02\\ (p=0.47)\end{tabular} &
  71.35 &
  70.19 &
  \begin{tabular}[c]{@{}c@{}}$\downarrow$ -1.16\\ (p=0.56)\end{tabular} &
  64.94 &
  69.35 &
  \begin{tabular}[c]{@{}c@{}}$\uparrow$ 4.40\\ (p=0.18)\end{tabular} \\ \midrule
\textbf{ConditionB} &
  67.86 &
  82.02 &
  \begin{tabular}[c]{@{}c@{}}$\uparrow$ 14.16\\ (p\textless{}0.001)\end{tabular} &
  72.22 &
  83.43 &
  \begin{tabular}[c]{@{}c@{}}$\uparrow$ 11.21\\ (p\textless{}0.05)\end{tabular} &
  64.58 &
  80.95 &
  \begin{tabular}[c]{@{}c@{}}$\uparrow$ 16.37\\ (p\textless{}0.01)\end{tabular} \\ \bottomrule
\end{tabular}%
}
\end{table*}

\begin{table*}[htp]
\centering
\caption{Detailed performance and team-gain metrics for all participants (All), domain experts (EXPs), and health professionals and students (NVs)}
\label{tab:results_teamgains_detail}
\resizebox{0.8\textwidth}{!}{%
\begin{tabular}{lccccccccc}\toprule
\multicolumn{1}{c}{\textbf{}} &
  \multicolumn{3}{c}{\textbf{All}} &
  \multicolumn{3}{c}{\textbf{Experts}} &
  \multicolumn{3}{c}{\textbf{Novices}} \\  \cmidrule(lr){2-4}\cmidrule(lr){5-7}\cmidrule(lr){8-10}
 \textbf{Metrics}&
  \textbf{CondA} &
  \textbf{CondB} &
  \textbf{\begin{tabular}[c]{@{}c@{}}$\Delta$ B-A \\ (P-Value)\end{tabular}} &
  \textbf{CondA} &
  \textbf{CondB} &
  \textbf{\begin{tabular}[c]{@{}c@{}}$\Delta$ B-A \\ (P-Value)\end{tabular}}  &
  \textbf{CondA} &
  \textbf{CondB} &
  \textbf{\begin{tabular}[c]{@{}c@{}}$\Delta$ B-A \\ (P-Value)\end{tabular}} \\ \midrule
Human &
  67.69 &
  67.86 &
  \begin{tabular}[c]{@{}c@{}}0.17\\ (p=0.86)\end{tabular} &
  71.35 &
  72.22 &
  \begin{tabular}[c]{@{}c@{}}0.87\\ (p=0.61)\end{tabular} &
  64.94 &
  64.58 &
  \begin{tabular}[c]{@{}c@{}}-0.36\\ (p=0.90)\end{tabular} \\ \midrule
AI &
  70.76 &
  73.00 &
  \begin{tabular}[c]{@{}c@{}}2.24\\ (p=0.14)\end{tabular} &
  70.58 &
  73.31 &
  \begin{tabular}[c]{@{}c@{}}2.73\\ (p=0.5)\end{tabular} &
  70.89 &
  72.77 &
  \begin{tabular}[c]{@{}c@{}}1.88\\ (p=1.00)\end{tabular} \\  \midrule
Human-AI Team &
  69.71 &
  82.02 &
  \begin{tabular}[c]{@{}c@{}}$\uparrow$ 12.31\\ (p\textless{}0.001)\end{tabular} &
  70.19 &
  83.43 &
  \begin{tabular}[c]{@{}c@{}}$\uparrow$ 13.25\\ (p\textless{}0.05)\end{tabular} &
  69.35 &
  80.95 &
  \begin{tabular}[c]{@{}c@{}}$\uparrow$ 11.61\\ (p\textless{}0.05)\end{tabular} \\ \midrule
\begin{tabular}[c]{@{}l@{}}Human-AI Team\_gain \\ vs Human\end{tabular} &
  2.02 &
  14.16 &
  \begin{tabular}[c]{@{}c@{}}$\uparrow$ 12.14\\ (p\textless{}0.01)\end{tabular} &
  -1.16 &
  11.21 &
  \begin{tabular}[c]{@{}c@{}}$\uparrow$ 12.37\\ (p\textless{}0.05)\end{tabular} &
  4.4 &
  16.37 &
  \begin{tabular}[c]{@{}c@{}}$\uparrow$ 11.96\\ (p\textless{}0.05)\end{tabular} \\ \midrule
\begin{tabular}[c]{@{}l@{}}Human-AI Team\_gain\\ vs\_AI\end{tabular} &
  -1.05 &
  9.01 &
  \begin{tabular}[c]{@{}c@{}}$\uparrow$ 10.07\\ (p\textless{}0.05)\end{tabular} &
  -0.4 &
  10.12 &
  \begin{tabular}[c]{@{}c@{}} 10.52\\ (p=0.20)\end{tabular} &
  -1.55 &
  8.18 &
  \begin{tabular}[c]{@{}c@{}}$\uparrow$ 9.73\\ (p\textless{}0.1)\end{tabular} \\ \midrule
Regret\_best &
  16.46 &
  4.80 &
  \begin{tabular}[c]{@{}c@{}}$\downarrow$ 11.66\\ (p\textless{}0.01)\end{tabular} &
  17.46 &
  3.87 &
  \begin{tabular}[c]{@{}c@{}}$\downarrow$ 13.59\\ (p\textless{}0.05)\end{tabular} &
  15.71 &
  5.51 &
  \begin{tabular}[c]{@{}c@{}}$\downarrow$ 10.21\\ (p\textless{}0.05)\end{tabular} \\ \midrule
Oracle\_best &
  86.17 &
  86.82 &
  \begin{tabular}[c]{@{}c@{}} 0.65\\ (p=0.63)\end{tabular} &
  87.65 &
  87.30 &
  \begin{tabular}[c]{@{}c@{}} 0.34\\ (p=1)\end{tabular} &
  85.06 &
  86.46 &
  \begin{tabular}[c]{@{}c@{}} 1.4\\ (p=0.68)\end{tabular} \\ \bottomrule
\end{tabular} 
}
\end{table*}

\begin{table*}[htp]
\centering
\caption{Detailed acceptance/rejection metrics capturing reliance on AI for all participants (All), domain experts (EXPs), and health professionals and students (NVs)}
\label{tab:results_reliance_detail}
\resizebox{0.8\textwidth}{!}{%
\begin{tabular}{lccccccccc} \toprule
\multicolumn{1}{c}{\textbf{}} &
  \multicolumn{3}{c}{\textbf{All}} &
  \multicolumn{3}{c}{\textbf{Experts}} &
  \multicolumn{3}{c}{\textbf{Novices}} \\ \cmidrule(lr){2-4}\cmidrule(lr){5-7}\cmidrule(lr){8-10}
\textbf{Metrics} &
  \textbf{CondA} &
  \textbf{CondB} &
  \textbf{\begin{tabular}[c]{@{}c@{}}$\Delta$ B-A\\ (P-Value)\end{tabular}} &
  \textbf{CondA} &
  \textbf{CondB} &
  \textbf{\begin{tabular}[c]{@{}c@{}}$\Delta$ B-A\\ (P-Value)\end{tabular}} &
  \textbf{CondA} &
  \textbf{CondB} &
  \textbf{\begin{tabular}[c]{@{}c@{}}$\Delta$ B-A\\ (P-Value)\end{tabular}} \\ \midrule
\begin{tabular}[c]{@{}l@{}}Accept\_on\_correct\end{tabular} &
  88.09 &
  95.24 &
  \begin{tabular}[c]{@{}c@{}} $\uparrow$ 7.15\\ (p\textless{}0.01)\end{tabular} &
  86.4 &
  98.89 &
  \begin{tabular}[c]{@{}c@{}}$\uparrow$ 12.49\\ (p\textless{}0.01)\end{tabular} &
  89.36 &
  92.5 &
  \begin{tabular}[c]{@{}c@{}}3.14\\ (p=0.34)\end{tabular} \\ \midrule
\begin{tabular}[c]{@{}l@{}}Reject\_on\_wrong\end{tabular} &
  41.61 &
  62.14 &
  \begin{tabular}[c]{@{}c@{}}$\uparrow$ 20.53\\ (p\textless{}0.05)\end{tabular} &
  42.33 &
  56.11 &
  \begin{tabular}[c]{@{}c@{}}13.78\\ (p=0.40)\end{tabular} &
  41.07 &
  66.67 &
  \begin{tabular}[c]{@{}c@{}}25.6\\ (p=0.1)\end{tabular} \\ \midrule
\begin{tabular}[c]{@{}l@{}}Reject\_on\_correct\end{tabular} &
  11.91 &
  4.76 &
  \begin{tabular}[c]{@{}c@{}}$\downarrow$ -7.15\\ (p\textless{}0.01)\end{tabular} &
  13.6 &
  1.11 &
  \begin{tabular}[c]{@{}c@{}}$\downarrow$ -12.49\\ (p\textless{}0.01)\end{tabular} &
  10.64 &
  7.5 &
  \begin{tabular}[c]{@{}c@{}}-3.14\\ (p=0.34)\end{tabular} \\ \midrule
\begin{tabular}[c]{@{}l@{}}Accept\_on\_wrong\end{tabular} &
  58.39 &
  37.86 &
  \begin{tabular}[c]{@{}c@{}}$\downarrow$ -20.53\\ (p\textless{}0.01)\end{tabular} &
  57.67 &
  43.89 &
  \begin{tabular}[c]{@{}c@{}}-13.78\\ (p=0.40)\end{tabular} &
  58.93 &
  33.33 &
  \begin{tabular}[c]{@{}c@{}}-25.6\\ (p=0.1)\end{tabular} \\ \midrule
AI\_help\_rate &
  13.76 &
  15.9 &
  \begin{tabular}[c]{@{}c@{}}2.14\\ (p=0.58)\end{tabular} &
  10.85 &
  15.08 &
  \begin{tabular}[c]{@{}c@{}}4.23\\ (p=0.62)\end{tabular} &
  15.95 &
  16.52 &
  \begin{tabular}[c]{@{}c@{}}0.57\\ (p=0.88)\end{tabular} \\ \midrule
AI\_harm\_rate &
  8.29 &
  3.7 &
  \begin{tabular}[c]{@{}c@{}}$\downarrow$ -4.59\\ (p\textless{}0.05)\end{tabular} &
  7.91 &
  3.87 &
  \begin{tabular}[c]{@{}c@{}}-4.04\\ (p=0.21)\end{tabular} &
  8.57 &
  3.57 &
  \begin{tabular}[c]{@{}c@{}}-5.00\\ (p=0.12)\end{tabular} \\ \midrule
Missed\_help\_rate &
  4.72 &
  3.06 &
  \begin{tabular}[c]{@{}c@{}}-1.66\\ (p=0.29)\end{tabular} &
  5.45 &
  0 &
  \begin{tabular}[c]{@{}c@{}}$\downarrow$ -5.45\\ (p\textless{}0.1)\end{tabular} &
  4.17 &
  5.36 &
  \begin{tabular}[c]{@{}c@{}}1.19\\ (p=0.76)\end{tabular} \\ \midrule
Correct\_ignore\_rate &
  7.12 &
  10.12 &
  \begin{tabular}[c]{@{}c@{}}3.00\\ (p=0.22)\end{tabular} &
  9.15 &
  10.12 &
  \begin{tabular}[c]{@{}c@{}}0.97\\ (p=0.64)\end{tabular} &
  5.6 &
  10.12 &
  \begin{tabular}[c]{@{}c@{}}4.52\\ (p=0.27)\end{tabular} \\ \bottomrule
\end{tabular}%
}
\end{table*}

\begin{table*}[htp]
\centering
\caption{Ratio of Changed, ChangedToRight, and ChangedToWrong rates after reviewing AI outputs (Human + AI) for all participants (All), domain experts (EXPs), and health professionals and students (NVs)}
\label{tab:results_changed_decisions_detail}
\resizebox{0.8\textwidth}{!}{%
\begin{tabular}{lccccccccc} \toprule
\multicolumn{1}{c}{\textbf{}} &
  \multicolumn{3}{c}{\textbf{All}} &
  \multicolumn{3}{c}{\textbf{Experts}} &
  \multicolumn{3}{c}{\textbf{Novices}} \\ \cmidrule(lr){2-4}\cmidrule(lr){5-7}\cmidrule(lr){8-10}

\textbf{Metrics} &
  \textbf{CondA} &
  \textbf{CondB} &
  \textbf{\begin{tabular}[c]{@{}c@{}}$\Delta$ B-A\\ (P-Value)\end{tabular}} &
  \textbf{CondA} &
  \textbf{CondB} &
  \textbf{\begin{tabular}[c]{@{}c@{}}$\Delta$ B-A\\ (P-Value)\end{tabular}} &
  \textbf{CondA} &
  \textbf{CondB} &
  \textbf{\begin{tabular}[c]{@{}c@{}}$\Delta$ B-A\\ (P-Value)\end{tabular}} \\ \midrule
Changed\_Rate &
  34.98 &
  26.66 &
  \begin{tabular}[c]{@{}c@{}}$\downarrow$-8.32\\ (p\textless{}0.1)\end{tabular} &
  32.41 &
  26.09 &
  \begin{tabular}[c]{@{}c@{}}-6.32\\ (p=0.41)\end{tabular} &
  36.9 &
  27.08 &
  \begin{tabular}[c]{@{}c@{}}-9.82\\ (p=0.10)\end{tabular} \\ \midrule
ChangedToRight\_Rate &
  14.74 &
  18.88 &
  \begin{tabular}[c]{@{}c@{}}4.14\\ (p=0.13)\end{tabular} &
  12.57 &
  17.46 &
  \begin{tabular}[c]{@{}c@{}}4.89\\ (p=0.34)\end{tabular} &
  16.37 &
  19.94 &
  \begin{tabular}[c]{@{}c@{}}3.57\\ (p=0.27)\end{tabular} \\ \midrule
ChangedToWrong\_rate &
  12.72 &
  4.72 &
  \begin{tabular}[c]{@{}c@{}}$\downarrow$ -8.00\\ (p\textless{}0.01)\end{tabular} &
  13.73 &
  6.25 &
  \begin{tabular}[c]{@{}c@{}}$\downarrow$ -7.48\\ (p\textless{}0.05)\end{tabular} &
  11.96 &
  3.57 &
  \begin{tabular}[c]{@{}c@{}}$\downarrow$ -8.39\\ (p\textless{}0.05)\end{tabular} \\ \bottomrule
\end{tabular}%
}
\end{table*}

\subsection{Phase 2: Interactive Refinement and Offline Adaptation}
\subsubsection{Feedback-to-update mechanisms.}
\textbf{Re-label feedback (dataset curation for offline adaptation).}
For each trial where a participant disagrees with the AI output and provides a corrected label,
we create a feedback instance $(x,\tilde{y})$, where $x$ denotes the original model input
(e.g., patient trial/window features) and $\tilde{y}$ is the participant-provided label.
We operationalize these instances as additional supervised training data by augmenting
the original training set:
\[
\mathcal{D}^{\mathrm{global}}_{c}=\mathcal{D}_{\mathrm{base}}\cup \mathcal{F}^{\mathrm{global}}_{c},
\]
where 
$\mathcal{F}^{\mathrm{global}}_{c}$ pools re-label feedback across patients under condition $c$.
We then perform offline adaptation by fine-tuning the same baseline checkpoint using supervised learning
on 
$\mathcal{D}^{\mathrm{global}}_{c}$ (global), and evaluate
on a disjoint held-out set to measure generalization beyond the corrected cases.

\emph{Quality control.} Before augmentation, we (i) exclude incomplete feedback (missing $\tilde{y}$),
(ii) deduplicate identical instances (same $x$ and $\tilde{y}$), and
(iii) if multiple feedback labels exist for the same case, we consolidate them via a predefined
conflict-resolution policy (e.g., majority vote), yielding one consolidated label per case.

\textbf{Rule-based feedback (explicit inference-time update).}
Participants can create or edit rules that map detectable conditions (feature/threshold patterns)
to corrective actions on the model output. 

We treat rules as \emph{patient-specific} by default (authored for a given patient), and evaluate
their transfer by applying the same rule layer to other patients' held-out data.

\subsubsection{Local vs.\ global updates and generalization checks.}
For each patient $i$ and condition $c\in\{A,B\}$, we construct and evaluate local and global updates. 

\textbf{Local updates.} use only feedback associated with patient $i$ under condition $c$:
\emph{local rule update} by applying the edited patient-specific rule for patient $i$.

\textbf{Global updates / transfer checks.}
We use two complementary notions of ``global'' depending on feedback type:
(i) \emph{global re-label model} by fine-tuning a single model on $\mathcal{D}^{\mathrm{global}}_{c}$, and
(ii) \emph{cross-patient rule transfer} by applying a patient's edited rule layer to held-out data
from other patients $j\neq i$, measuring spillover effects beyond the authoring patient.

\paragraph{Refinement-performance metrics.}
Let $\mathrm{Perf}(\cdot)$ denote model performance (percentage; higher is better) on the held-out set.
We report improvement scores relative to the baseline:
\[
\Delta \mathrm{Perf}^{\mathrm{local}}_{i,c}
=
\mathrm{Perf}(\text{local update}_{i,c})
-
\mathrm{Perf}(\text{baseline})
\]
\[
\Delta \mathrm{Perf}^{\mathrm{global}}_{c}
=
\mathrm{Perf}(\text{global update}_{c})
-
\mathrm{Perf}(\text{baseline}).
\]

\begin{table*}[t]
\centering
\caption{Phase 2 refinement performance. We report improvement over the baseline model on a held-out evaluation set for local (participant-specific) and global (across-patients) updates. $p$ from paired Wilcoxon signed-rank tests across participants.}
\label{tab:phase2_refinement}
\small
\begin{tabular}{lccc c ccc c ccc c}
\toprule
 & \multicolumn{4}{c}{All (Experts+Novices)} & \multicolumn{4}{c}{Experts} & \multicolumn{4}{c}{Novices}\\
\cmidrule(lr){2-5}\cmidrule(lr){6-9}\cmidrule(lr){10-13}
Metric & A & B & $\Delta$B--A & $p$ & A & B & $\Delta$B--A & $p$ & A & B & $\Delta$B--A & $p$\\
\midrule
Local update $\Delta$Perf (overall)  & 11.50 & 36.38 & 24.89 & <0.001 & 11.20 & 33.89 & 22.69 & <0.001 & 11.92 & 38.88 & 26.96 & <0.05\\
Global update $\Delta$Perf (overall) & 2.38  & -5.05 & -7.43 & <0.001 & 5.58  & -2.95 & -8.52 & <0.001 & 0.00  & -6.46 & -6.46 & <0.01\\
\bottomrule
\end{tabular}
\end{table*}


\end{document}